\documentclass[a4paper,11pt]{article}
\pdfoutput=1
\usepackage{amssymb,amsmath,amsfonts,makeidx,placeins,pbox,multirow}
\usepackage{graphicx,rotate,subcaption,color,slashed,cite,caption,epstopdf,verbatim}
\usepackage{longtable,tabu}
\usepackage{array}
\newcolumntype{P}[1]{>{\centering\arraybackslash}p{#1}}
\usepackage[colorlinks=true,
            linkcolor=magenta,
            urlcolor=blue,
            citecolor=blue]{hyperref}

\numberwithin{equation}{section}
\numberwithin{figure}{section}
\numberwithin{table}{section}

\newenvironment{rcases}
  {\left.\begin{aligned}}
  {\end{aligned}~~\right\rbrace}

\evensidemargin=\oddsidemargin
\begin{document}
%\maketitle

\begin{center}

{\Large \bf Constraining Composite Higgs Models using LHC data} \\
\vspace*{0.5cm} {\sf Avik
  Banerjee~$^{a,}$\footnote{avik.banerjeesinp@saha.ac.in}, ~Gautam
  Bhattacharyya~$^{a,}$\footnote{gautam.bhattacharyya@saha.ac.in}, ~Nilanjana Kumar~$^{a,}$\footnote{nilanjana.kumar@saha.ac.in}, ~Tirtha
  Sankar Ray~$^{b,}$\footnote{tirthasankar.ray@gmail.com}} \\
\vspace{10pt} {\small } $^{a)}$ {\em Saha Institute of Nuclear
    Physics, HBNI, 1/AF Bidhan Nagar, Kolkata 700064, India}
\\
\vspace{3pt} {\small } $^{b)}${\em Department of Physics and Centre
  for Theoretical Studies, Indian Institute of Technology Kharagpur,
  Kharagpur 721302, India}
\normalsize
\end{center}

% % % % % % % % % % % % % % % % Abstract % % % % % % % % % % % % % % % % % 

\begin{abstract}

We systematically study the modifications in the couplings of the
Higgs boson, when identified as a pseudo Nambu-Goldstone boson of a
strong sector, in the light of LHC Run 1 and Run 2 data. For the
minimal coset $\rm SO(5)/SO(4)$ of the strong sector, we focus on
scenarios where the standard model left- and right-handed fermions
(specifically, the top and bottom quarks) are either in $\bf 5$ or in
the symmetric $\bf 14$ representation of $\rm SO(5)$. Going beyond the
minimal $\bf 5_L-5_R$ representation, to what we call here the
`extended' models, we observe that it is possible to construct more
than one invariant in the Yukawa sector.  In such models, the Yukawa
couplings of the 125 GeV Higgs boson undergo nontrivial modifications.
The pattern of such modifications can be encoded in a generic
phenomenological Lagrangian which applies to a wide class of such
models.  We show that the presence of more than one Yukawa invariant
allows the gauge and Yukawa coupling modifiers to be decorrelated in
the `extended' models, and this decorrelation leads to a relaxation of
the bound on the compositeness scale ($f \geq640$ GeV at $95 \%$ CL,~
as compared to $f \geq 1$ TeV for the minimal $\bf 5_L-5_R$
representation model).  We also study the Yukawa coupling
modifications in the context of the next-to-minimal strong sector
coset $\rm SO(6)/SO(5)$ for fermion-embedding up to representations of
dimension $\bf 20$.  While quantifying our observations, we have
performed a detailed $\chi^2$ fit using the ATLAS and CMS combined Run
1 and available Run 2 data.

\end{abstract}

% % % % % % % % % % % % % % % % % % % % % % % % % % % % % % % % % % % % % %

\bigskip

% % % % % % % % % % % % % % % Main Document% % % % % % % % % % % % % % % % 

\section{Introduction}
\label{intro}

With increasing precision in measurements of the Higgs boson
properties at the Large Hadron Collider (LHC), the possibility that
the Higgs may be a composite object
\cite{Kaplan:1983fs,Dugan:1984hq,Contino:2010rs,Panico:2015jxa,Csaki:2016kln}
can be put to stringent tests. In this context, the scenarios where
the Higgs is identified as a pseudo Nambu-Goldstone boson (pNGB) of a
strongly interacting sector are of special interest.  This has
received considerable attention following its identification as a
holographic dual of 5d gauge-Higgs unification models
\cite{Contino:2003ve,Contino:2006nn,Csaki:2017eio,Espriu:2017mlq}. In
this paper, however, we stick to an effective 4d scenario, and do not
comment on possible UV completion of such models. The approximate
shift-symmetry of the pNGBs can screen the weak scale from physics
beyond the compositeness scale ($f\sim \mathcal{O}(\rm TeV)$).  This
provides a well-motivated framework for natural electroweak symmetry
breaking.

The direct signatures of these models at the LHC could be the
appearance of additional resonances of the strong sector
\cite{DeSimone:2012fs,Thamm:2015zwa, Azatov:2015xqa,
  Araque:2015cna,Aaboud:2017qpr,Sirunyan:2017jin,Sirunyan:2017usq,Chala:2017xgc,Yepes:2017pjr}. However
taking cue from non-observation of these resonances, attempts have
been made to push up the resonance masses while keeping the theory
still natural \cite{Pappadopulo:2013vca,Croon:2015wba,
  Banerjee:2017qod,Csaki:2017jby,Chacko:2005pe}.  The other inevitable
and testable features of these models are deviations of the Higgs
couplings compared to their standard model (SM) predictions.  One of
the consequences of compositeness is that the couplings are replaced
by form factors which are momentum dependent. However, it is difficult
to test this momentum dependence at the LHC.  Nevertheless, the
nonlinearity of the pNGB dynamics provides a finite shift in the Higgs
couplings measurable in the precision era of the LHC.  In this paper
we make a systematic study of the pattern and constraints on such
modifications that arise in a general class of composite Higgs models.

We categorize the scenarios considered under three major heads:
\begin{itemize}
\item {\em Minimal model}: Coset $\rm SO(5)/SO(4)$, with both the
  left- and right-handed fermions kept in the fundamental $\bf 5$ of
  $\rm SO(5)$, represented in literature as
  MCHM$_{5_L-5_R}$
  \cite{Agashe:2004rs,Panico:2011pw,Matsedonskyi:2012ym,Marzocca:2012zn,Pomarol:2012qf,Carena:2014ria}.

\item {\em Extended models}: Coset $\rm SO(5)/SO(4)$, with at least
  one of the left- or right-handed fermions kept in the symmetric
  $\bf14$ of $\rm SO(5)$. They are denoted in literature as
  MCHM$_{14_L-14_R}$, MCHM$_{14_L-5_R}$, and MCHM$_{5_L-14_R}$
  \cite{Panico:2012uw,Montull:2013mla,Carena:2014ria,Carmona:2014iwa,Kanemura:2016tan,Gavela:2016vte,Liu:2017dsz}.

\item {\em Next-to-minimal models}: Coset $\rm SO(6)/SO(5)$, denoted
  as NMCHM, with different choices of representation up to dimension
  $\bf 20$ \cite{Gripaios:2009pe,Redi:2012ha, Barnard:2013zea,
    Serra:2015xfa,Low:2015qep,Cai:2015bss,Arbey:2015exa,Niehoff:2016zso,Banerjee:2017qod,Sanz:2017tco}.

\end{itemize}

The couplings of the pNGB Higgs with the weak gauge bosons ($VVh$) are
usually suppressed in a general class of composite models. The
parameter $\xi \equiv \mathnormal{v}^2/f^2\ll 1$, where $v=246$ GeV is the
electroweak vacuum expectation value (vev), controls this
suppression. The Yukawa couplings are generated through a mixing
between the elementary fermions and the operators of the strong
sector. Once the strong sector is integrated out the effective
Higgs-fermion interaction term looks like \cite{Giudice:2007fh,
  Contino:2013kra},
\begin{equation}
\label{intro_2}
{\mathcal{L}}_{eff} \propto \bar{f}_LHf_R\mathcal{F}\left(\frac{H^\dagger H}{f^2}\right)~,
\end{equation}
where $\mathcal{F}(H^\dagger H/f^2)$ is a function of the $\rm
SU(2)_L$ doublet Higgs field ($H$). The contributions from the higher
dimensional operators with independent coefficients, added to the SM
dimension-4 Yukawa term, give rise to a modification in the couplings
of the Higgs with the fermions ($f\bar{f}h$), see also
\cite{Hashimoto:2017jvc,Jana:2017hqg} in a different context.  In the
minimal model, the SM fermions couple to only one operator of the
strong sector. As a result the modification of the couplings depends
on only one free parameter $\xi$.  The other parameters in the
effective Lagrangian are fixed from the requirement of reproducing the
corresponding SM fermion mass.  Therefore, $f\bar{f}h$ and $VVh$
couplings get highly correlated, and stringent constraints on $f$
emerge \cite{Cacciapaglia:2012wb,Falkowski:2013dza} from the
increasingly precise measurements of Higgs production and decays at
LHC.  In the extended models, owing to the presence of more than one
invariant in the Yukawa sector with different coefficients, the
correlation between $f\bar{f}h$ and $VVh$ modifiers is weakened, and
we observe a possible relaxation of the bound on $f$. This happens in
certain regions of the parameter space where a possible enhancement in
$f\bar{f}h$ vertex can partially offset the suppression in $VVh$
coupling. Additionally, the extended models, carrying more than one
invariant in the Yukawa sector, have the distinct advantage of being
free from `double tuning' \cite{Panico:2012uw}\footnote{$\Delta =
  1/\xi$ is a measure of minimal tuning in any composite Higgs
  model. On top of this, an additional tuning, dubbed `double tuning',
  arises in scenarios (e.g.~MCHM$_{5_L-5_R}$) where the coefficients
  of the quadratic and quartic terms in the potential are not in the
  same order of the elementary-composite mixing parameter. This can be
  avoided when either the fermion kinetic and/or the Yukawa terms
  contain at least two invariants.}.

In this paper, we first concentrate on a systematic and comparative
study of various possibilities of Higgs coupling modifications in the
context of the extended models\footnote{We do not consider
  representation {\bf 10} of SO(5) because it does not lead to more
  than one Yukawa invariant keeping a discrete parity that protects
  the $Zb\bar b$ vertex \cite{Carena:2014ria,Azatov:2013ura}. Note
  that the choice MCHM$_{14_L-1_R}$, where $t_R$ can be fully
  composite, involves minimal tuning as compared to the double tuned
  MCHM$_{5_L-5_R}$ \cite{Panico:2012uw}. However, we do not consider this choice because it
  contains a single Yukawa invariant.}. For each such possibilities,
we construct one-loop Coleman-Weinberg Higgs potential
\cite{Coleman:1973jx}, and identify regions of parameters space where
the top mass, Higgs mass and the electroweak vev are reproduced.  Next
we consider the next-to-minimal model which contains a SM
scalar-singlet ($\eta$) apart from the Higgs doublet. Their mixing can
significantly modify the observed Higgs boson couplings. In this
context also, we survey different fermionic representations and
calculate the corresponding modifications to Yukawa couplings.

We then construct an effective phenomenological Lagrangian whose
parameters capture the coupling modifications of a general class of
models mentioned earlier.  The explicit connection between the
coefficients of the Lagrangian and the parameters of the specific
models is specified on a case-by-case basis.  We perform a $\chi^2$
analysis with the ATLAS and CMS combined Run 1
\cite{Khachatryan:2016vau} and available Run 2 data
\cite{ATLAS-CONF-2016-112,ATLAS-CONF-2017-045,ATLAS-CONF-2017-077,ATLAS-CONF-2017-043,Aaboud:2017xsd,ATLAS-CONF-2017-076,CMS-PAS-HIG-16-021,CMS-PAS-HIG-16-040,Sirunyan:2017exp,Sirunyan:2017khh,Sirunyan:2017elk,CMS-PAS-HIG-17-003,CMS-PAS-HIG-17-004}
to estimate a bound on $f$ in the extended models and compare it with
that of the minimal model. In the context of the next-to-minimal
model, we provide an estimate of the amount of doublet-singlet scalar
mixing allowed by the current data.

The rest of the paper is organized as follows. In Section \ref{models}
we review the consequences of various fermionic embedding for 
$\rm SO(5)/SO(4)$ as well as $\rm SO(6)/SO(5)$ cosets. 
 In
Section \ref{eff_lag} we present a phenomenological Lagrangian that
captures the generic features of a wide class of models in terms of the
Higgs coupling modifiers. Following this parametrization we perform a fit
to the existing data using the $\chi^2$ minimization technique in Section
\ref{pheno}. Finally, we draw our conclusions in Section \ref{conc}.

% % % % % % % % % % % % % % % % % % % % % % % % % % % % % % % % % % % % % %

\section{Composite Models and Modified Yukawa Couplings}
\label{models}

In this section we consider different representations for fermions in $\rm SO(5)/SO(4)$ and $\rm SO(6)/SO(5)$ cosets and work out the modifications in the top quark Yukawa coupling in a systematic manner. 

% % % % % % % % % % % % % % % % % % % % % % % % % % % % % % % % % % % % % %

\subsection{$\rm\bf SO(5)/SO(4)$ Coset}
\label{minimal}

As long as the coset is $\rm SO(5)/SO(4)$, the modification in $VVh$ coupling is solely determined by $\xi$, as
\begin{equation}
\label{model_1}
k_{VVh}=\frac{g_{VVh}}{(g_{VVh})_{SM}}=\sqrt{1-\xi}\simeq 1-\frac{1}{2}\xi~.
\end{equation}

The number of Yukawa invariants, on the other hand, depends on the
representations in which $t_L$ and $t_R$ are embedded. We write down
the relevant invariants using the pNGB representation 
$\Sigma=\left(0,0,0,h/f,\sqrt{1-h^2/f^2}\right)^T$ in the unitary
gauge :
\begin{itemize}

\item $t_L$ and $t_R$ in $\bf 5$ ($\rm MCHM_{5_L-5_R}$):~~~~~$(\overline{Q}_{t_L}^{5}.\Sigma)(\Sigma^T.T_{t_R}^{5})$~,

\item $t_L$ and $t_R$ in $\bf 14$ ($\rm MCHM_{14_L-14_R}$):~~$\Sigma^T.\overline{Q}_{t_L}^{14}.T_{t_R}^{14}.\Sigma$~,~ $(\Sigma^T.\overline{Q}_{t_L}^{14}.\Sigma)(\Sigma^T.T_{t_R}^{14}.\Sigma)$~,

\item $t_L$ in $\bf 14$, $t_R$ in $\bf 5$ ($\rm MCHM_{14_L-5_R}$):~~$\Sigma^T.\overline{Q}_{t_L}^{14}.T_{t_R}^{5}$~,~$(\Sigma^T.\overline{Q}_{t_L}^{14}.\Sigma)(\Sigma^T.T_{t_R}^{5})$~,

\item $t_L$ in $\bf 5$, $t_R$ in $\bf 14$ ($\rm MCHM_{5_L-14_R}$):~~$\overline{Q}_{t_L}^{5}.T_{t_R}^{14}.\Sigma$~,~$(\overline{Q}_{t_L}^{5}.\Sigma)(\Sigma^T.T_{t_R}^{14}.\Sigma)$~.
\end{itemize} 
Above, $Q_{t_L}$ and $T_{t_R}$ contain $t_L$ and $t_R$ as incomplete $\rm SO(5)$ multiplets, respectively (see Appendix \ref{mchm_embeddings}).
The most general Lagrangian involving the top quark can be written as
\begin{equation}
\label{exp_model_2}
\mathcal{L}=\overline{t}_L\slashed q\Pi_{t_L}(q,h)t_L+\overline{t}_R\slashed q\Pi_{t_R}(q,h)t_R+\overline{t}_L\Pi_{t_Lt_R}(q,h)t_R+\rm h.c.
\end{equation} 

\begin{table} 
\begin{longtable}{|c|c|c|c|}
\hline

\rule[-2ex]{0pt}{5.5ex} Models & $\Pi_{t_L}(q,h)$ & $\Pi_{t_R}(q,h)$ & $\Pi_{t_Lt_R}(q,h)$ \\ 

\hline 

\rule[-2ex]{0pt}{5.5ex} $\rm MCHM_{5_L-5_R}$ & $\Pi^L_0+\Pi^L_1 \frac{h^2}{f^2}$ & $\Pi^R_0+\Pi^R_1 \frac{h^2}{f^2}$ & $\Pi^{LR}_1\frac{h}{f}\sqrt{1-\frac{h^2}{f^2}}$\\

\hline 

\rule[-2ex]{0pt}{5.5ex} $\rm MCHM_{14_L-14_R}$ & $\Pi^L_0+\Pi^L_1 \frac{h^2}{f^2}+\Pi^L_2 \frac{h^4}{f^4}$ & $\Pi^R_0+\Pi^R_1 \frac{h^2}{f^2}+\Pi^R_2 \frac{h^4}{f^4}$ & $\frac{h}{f}\sqrt{1-\frac{h^2}{f^2}}\left(\Pi^{LR}_1+\Pi^{LR}_2 \frac{h^2}{f^2}\right)$\\

\hline 

\rule[-2ex]{0pt}{5.5ex} $\rm MCHM_{14_L-5_R}$ & $\Pi^L_0+\Pi^L_1 \frac{h^2}{f^2}+\Pi^L_2 \frac{h^4}{f^4}$ & $\Pi^R_0+\Pi^R_1 \frac{h^2}{f^2}$ & $\frac{h}{f}\left(\Pi^{LR}_1+\Pi^{LR}_2 \frac{h^2}{f^2}\right)$\\

\hline 

\rule[-2ex]{0pt}{5.5ex} $\rm MCHM_{5_L-14_R}$ & $\Pi^L_0+\Pi^L_1 \frac{h^2}{f^2}$ & $\Pi^R_0+\Pi^R_1 \frac{h^2}{f^2}+\Pi^R_2 \frac{h^4}{f^4}$ & $\frac{h}{f}\left(\Pi^{LR}_1+\Pi^{LR}_2 \frac{h^2}{f^2}\right)$\\

\hline

\caption{\small\it List of $\Pi$-functions, defined in Eq.~\eqref{exp_model_2} for different representations.}
\label{exp_model_table1}
\end{longtable}
\end{table}

The dependence on the strong sector dynamics is encoded inside the momentum dependent $\Pi$-functions. In Table~\ref{exp_model_table1}, we show the explicit forms of those functions for various representations in terms of the Higgs field with coefficients $\Pi^{L,R,LR}_{0,1,2}(q)$. The expressions for the latter in terms of the masses ($m_i$) and decay constants ($F^{L,R}_i$) of the strong sector resonances are given in the Appendix \ref{form factors} for the extended models, namely, $\rm MCHM_{14_L-14_R}$, $\rm MCHM_{14_L-5_R}$ and $\rm MCHM_{5_L-14_R}$ respectively. The mass of the top quark  and the modification of top Yukawa can be calculated from Eq.~\eqref{exp_model_2} as
\begin{equation}
\label{exp_model_3}
m_t=\frac{|\Pi_{t_Lt_R}(q,h)|}{\sqrt{\Pi_{t_L}(q,h)\Pi_{t_R}(q,h)}}\bigg|_{q\rightarrow 0,~h\rightarrow v}~,~~k_{t\overline{t}h}=\frac{y_{t\overline{t}h}}{\left(y_{t\overline{t}h}\right)_{SM}}=\frac{1}{\left(y_{t\overline{t}h}\right)_{SM}}\left(1-\frac{1}{2}\xi\right)\frac{\partial m_t}{\partial v}.
\end{equation}
In the second equality of Eq.~\eqref{exp_model_3}, the factor $\left(1-\frac{1}{2}\xi\right)$ arises due to canonical normalization of the Higgs field. As argued in \cite{Azatov:2011qy, Montull:2013mla}, the top quark contribution to the effective gluon-gluon-Higgs ($ggh$) coupling in composite Higgs models is independent of the wave function renormalization effects of the top quark due to cancellation with resonance loops. This would imply a deviation in effective $ggh$ coupling compared to the effective $t\overline{t}h$ coupling. The modification of the effective $ggh$ coupling can be expressed as
\begin{equation}
\label{exp_model_4}
k_{ggh}^{(t)}=\frac{c_{ggh}}{\left(c_{ggh}\right)_{SM}}=\frac{1}{\left(c_{ggh}\right)_{SM}}\left(1-\frac{1}{2}\xi\right)\frac{\partial \log|\Pi_{t_Lt_R}(q,h)|}{\partial h}\bigg|_{q\rightarrow 0,~h\rightarrow v}.
\end{equation}
The one-loop Coleman-Weinberg Higgs potential receives largest contribution from the top quark, as
\begin{equation}
\label{exp_model_5}
V_{\rm eff}=-2N_c\int \frac{d^4q}{(2\pi)^4}\log\left(q^2\Pi_{t_L}\Pi_{t_R}+|\Pi_{t_Lt_R}|^2\right)\simeq-\alpha \frac{h^2}{f^2}+\beta\frac{h^4}{f^4}.
\end{equation} 
The coefficients $\alpha$ and $\beta$ above are integrals over the form factors. A similar contribution to $\alpha$ arises from gauge boson loops  with opposite sign (largest contribution from $\rm SU(2)_L $ gauge bosons), parametrized as \cite{Frigerio:2011zg,Barnard:2014tla}
\begin{equation}
\label{exp_model_6}
\alpha_{g}\simeq -c_{g}\frac{1}{16\pi^2}\frac{9}{2}g^2g_{\rho}^2f^4,
\end{equation} 
where $c_g$ is an $\mathcal{O}(1)$ positive constant absorbing the details of the integration, $g$ is the $\rm SU(2)_L$ gauge coupling, and $g_\rho$ corresponds to that of strong sector spin-1 resonances. The gauge contribution to $\beta$ is numerically small. To calculate the top-induced contribution to $\alpha$ and $\beta$, we use certain parametrization of the momentum dependent form factors based on scaling arguments. The decay constants and the top-partner masses are parametrized as
\begin{equation}
\label{exp_model_7}
F^{L,R}_{i}=\lambda^{L,R}_{i} f~,~~~~m_{i}=g_{i} f,
\end{equation}
where $\lambda^{L,R}_{i}$ are dimensionless constants and $g_{i}$ denote strong couplings. 
In the present analysis, we keep $\left|\lambda_i\right|/g_i<1$. The strong sector coupling strengths are kept well within the perturbative limits, \emph{i.e.} $1<g_i<2\pi$. 
\begin{figure}[]
\centering
\begin{subfigure}[t]{0.442\textwidth}
\centering
\includegraphics[width=\linewidth]{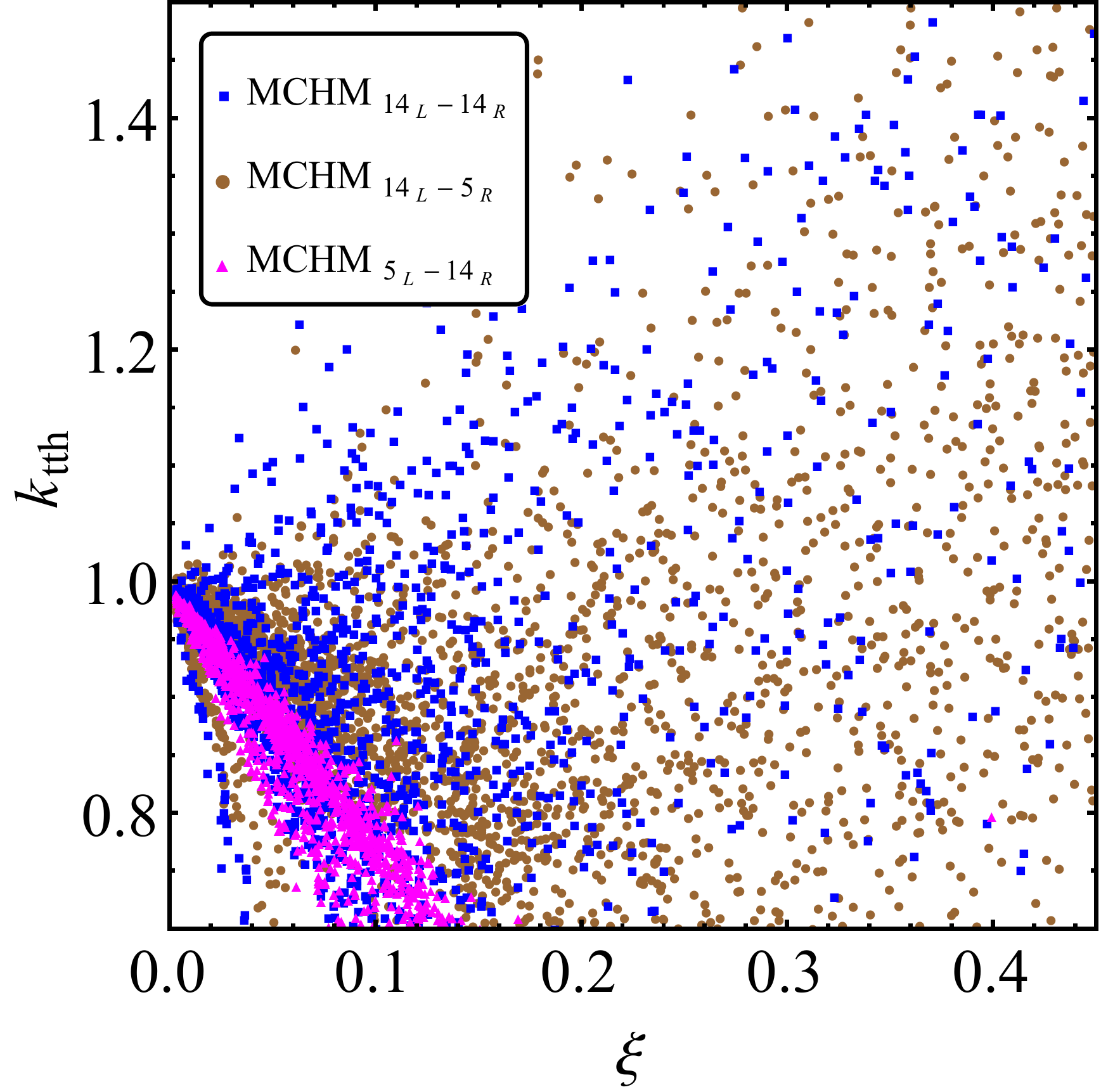}
%\caption{}
\end{subfigure}
~
\begin{subfigure}[t]{0.45\textwidth}
\centering
\includegraphics[width=\linewidth]{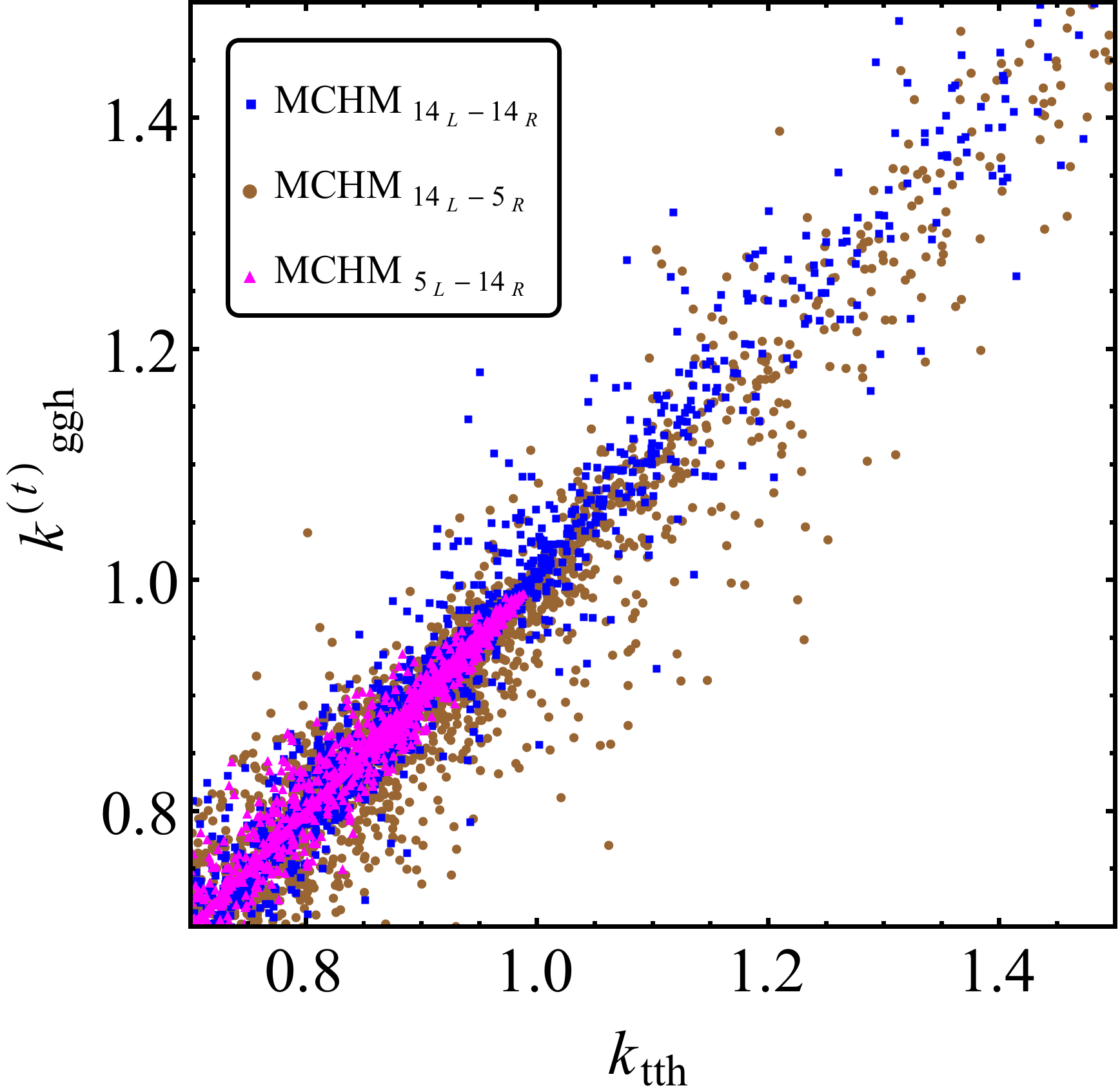}
%\caption{}
\end{subfigure}
\caption{\small\it Results from the numerical analysis for $\rm MCHM_{14_L-14_ R}$ (blue), $\rm MCHM_{14_L-5_R}$ (brown) and $\rm MCHM_{5_L-14_R}$ (magenta) are shown. In the left panel $k_{t\overline{t}h}$ is plotted against $\xi$ while at the right panel we show the correlation between $k_{ggh}^{(t)}$ and $k_{t\overline{t}h}$. While generating the model points we vary the strong couplings $g_i$ and $g_\rho$ in the range $[1,2\pi]$ and $\lambda^{L,R}_i/g_i$ within $[-1,1]$. All the points shown in the plots satisfy the phenomenological constraints given in Eqs.~\eqref{exp_model_9}.}
\label{numerical_fig1}
\end{figure}
Regarding the integrals over the form factors in fermionic sector, the
loop factors and the dimensionful variables are shown explicitly.
Some group theoretic factors also emerge due to the decomposition of
the SO(5) resonances in terms of SO(4) multiplets.  We assume that sufficient number of
resonances with coupling strengths $g_i$ saturate the form factors,
rendering the integrals finite.  As an illustration, we show one of
the integrals involved in the Higgs potential, as parametrized in
\cite{Contino:2010rs,Frigerio:2011zg,Barnard:2014tla}
\begin{eqnarray}
\label{exp_model_8}
\int \frac{d^4q}{(2\pi)^4}\left(\frac{\Pi^{L,R}_{1,2}(q)}{\Pi^{L,R}_0(q)}\right)^n &\simeq& c^{(n)}_{1,2}\frac{1}{16\pi^2}\left(\frac{\Pi^{L,R}_{1,2}(0)}{\Pi^{L,R}_0(0)}\right)^ng_i^4f^4~,~~~n=1,2, %\\
\end{eqnarray}
where $c^{(n)}_{1,2}$ are $\mathcal{O}(1)$ numbers and the forms
factors are displayed in Appendix \ref{form factors}.
Finally we use the following phenomenological constraints to generate the allowed parameter space:
\begin{eqnarray}
\label{exp_model_9}
\nonumber
169~\mathrm{GeV} < m_t< 176~\mathrm{GeV},~~v=246~ \mathrm{GeV},\\~~123~\mathrm{GeV} < m_h< 127~\mathrm{GeV},~~1~\mathrm{TeV} < m_i=g_if < 2\pi f ~.
\end{eqnarray}
We present the results of our numerical analysis in Fig.~\ref{numerical_fig1}. Depending on the embedding of the top quark the value of $k_{t\overline{t}h}$ varies.
Interestingly, for $\rm MCHM_{14_L-14_R}$ and $\rm MCHM_{14_L-5_R}$ we get an enhancement in top Yukawa coupling compared to its SM value ($k_{t\overline{t}h}>1$), for a large number of model points. On the other hand, for $\rm MCHM_{5_L-14_R}$ the top Yukawa is always suppressed. This is linked to the relative sign between the coefficients of the two Yukawa invariants. In Fig.~\ref{numerical_fig1} (right panel) we show the variation of $k_{ggh}^{(t)}$ with $k_{t\overline{t}h}$, and one observes that the two quantities are almost equal for all model points. This implies that the numerical impact of the wave function renormalization of the top quark is very small.   

% % % % % % % % % % % % % % % % % % % % % % % % % % % % % % % % % % % % % %

\subsection{$\rm \bf SO(6)/SO(5)$ Coset}
\label{nmchm}

The next-to-minimal model, with $\rm SO(6)/SO(5)$ coset includes a real singlet scalar ($\eta$) along with the usual Higgs doublet. Quite a few interesting features emerge in this case, depending on whether $\eta$ acquires a vev \cite{Niehoff:2016zso, Banerjee:2017qod, Sanz:2017tco} or not \cite{Frigerio:2012uc,Marzocca:2014msa,Fonseca:2015gva,Kim:2016jbz}. 
In the present section we discuss the effect of the $\eta$-vev and consequently the doublet-singlet scalar mixing on the Yukawa couplings. Here we follow the convention and notation as presented in \cite{Banerjee:2017qod}. 

In this case the structure of the Lagrangian involving the top quark is similar to Eq.~\eqref{exp_model_2}, with the exception that the $\Pi$-functions are dependent both on $h$ and $\eta$, as shown in Appendix \ref{nmchm_struc}, for different representations. 
Although compared to $\rm SO(5)/SO(4)$ coset, more possibilities of embedding $t_L$ and $t_R$ in different $\rm SO(6)$ multiplets exist, we stick to the choices shown in Appendix \ref{nmchm_embeddings} only. 
The Lagrangian, in terms of the canonically normalized quantum fields $(h_n, \eta_n)$, upon electroweak symmetry breaking, can be written as
\begin{equation}
\label{nmchm_1}
\mathcal{L}\supset m_t\overline{t}t+k_{t\overline{t}h_n} \left(\frac{m_t}{\mathnormal{v}}\right)h_n\overline{t}t+k_{t\overline{t}\eta_n} \left(\frac{m_t}{\mathnormal{v}}\right)\eta_n\overline{t}t~.
\end{equation}
\begin{figure}[t]
\centering
\begin{subfigure}[t]{0.45\textwidth}
\includegraphics[width=\linewidth]{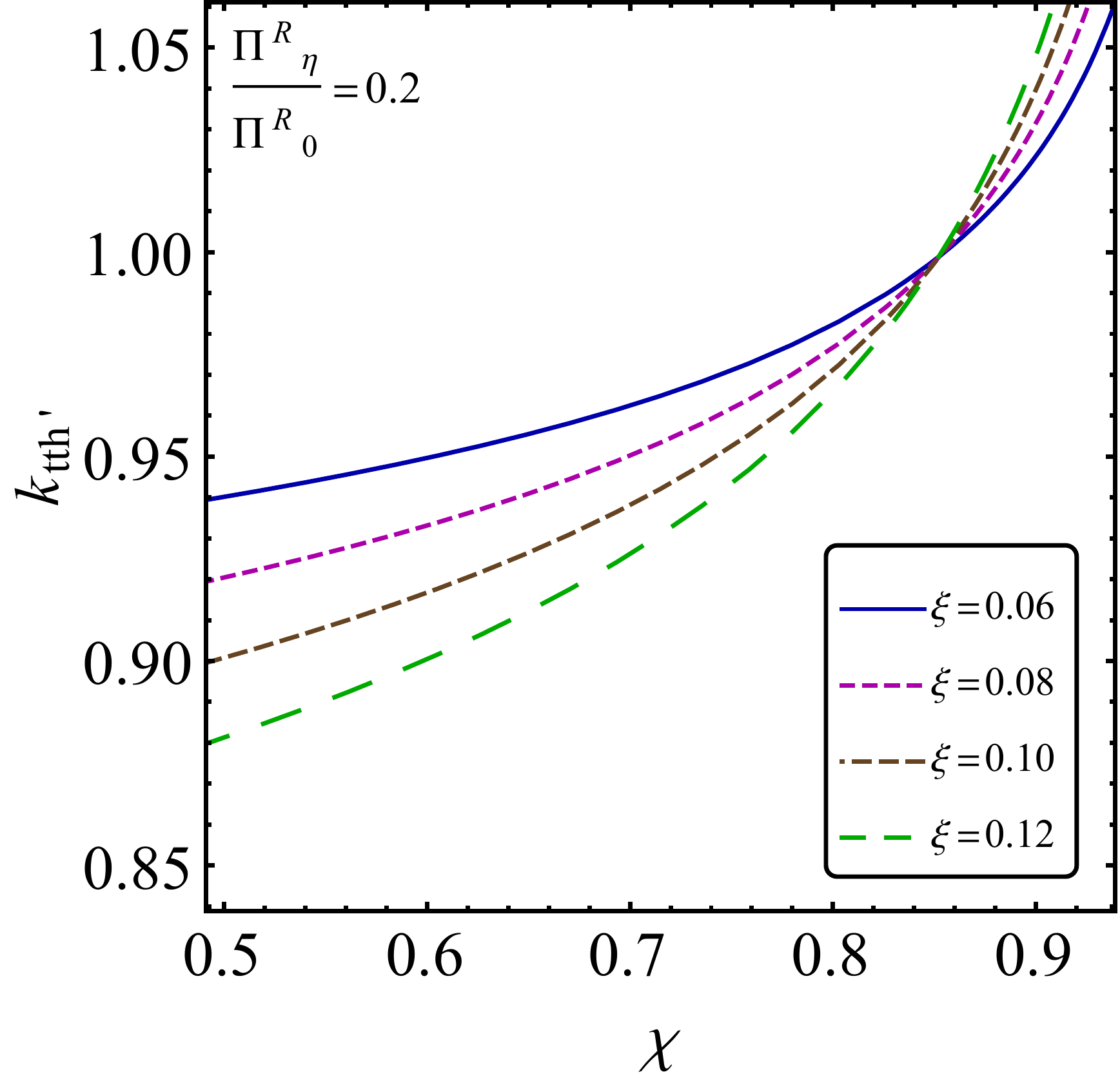}
%\caption{}
\end{subfigure} %
~
\begin{subfigure}[t]{0.448\textwidth}
\centering
\includegraphics[width=\linewidth]{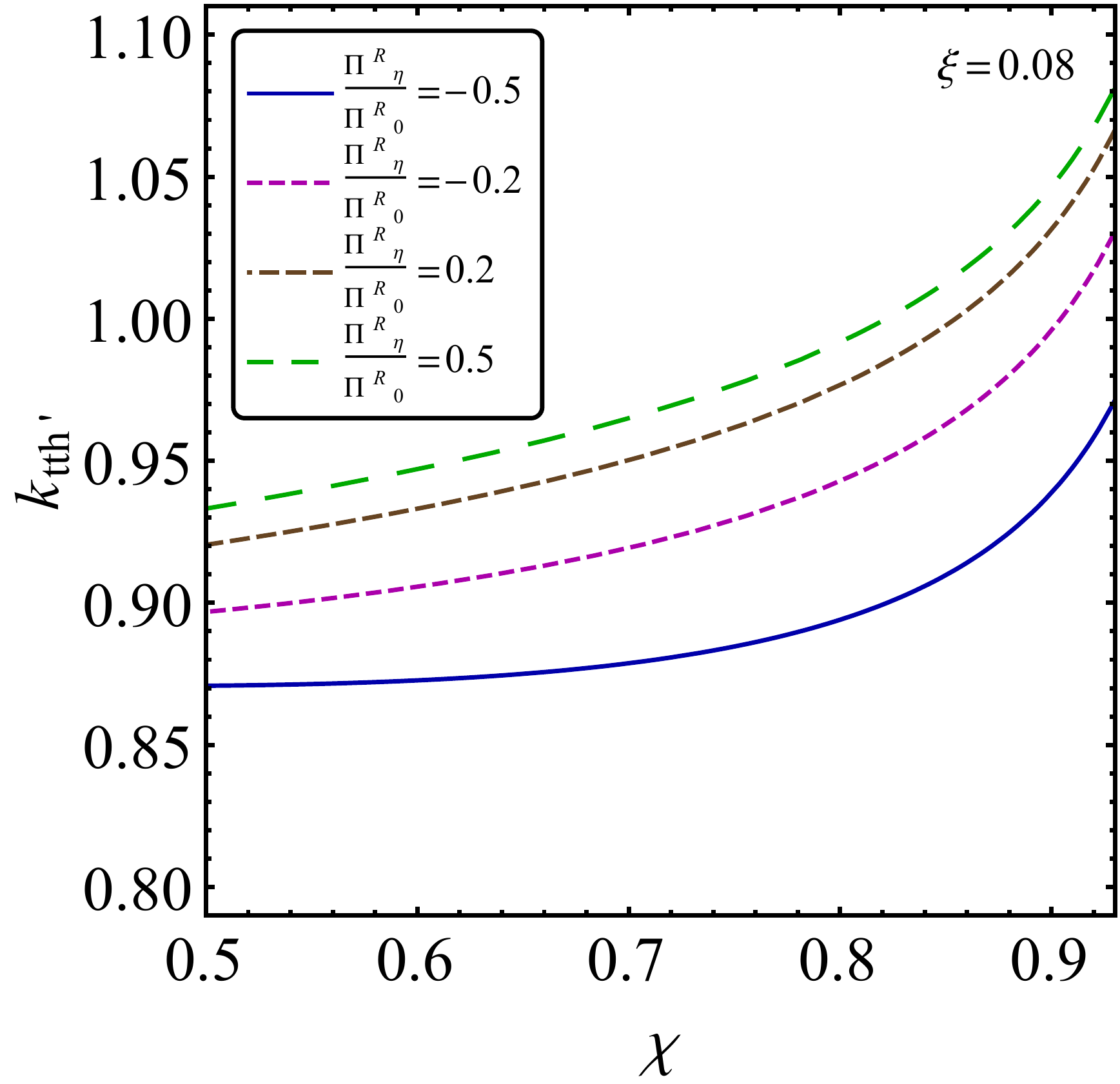}
%\caption{}
\end{subfigure} %
\caption{\small\it The variation of $k_{t\overline{t}h^\prime}$ with $\chi=\langle\eta\rangle^2/f^2$ is shown. In the left panel, for illustration, we keep the ratio $\Pi^R_\eta/\Pi^R_0$ (see Appendix~\ref{nmchm_struc}) fixed at 0.2, and plot the curves for different values of $\xi$. In the right panel we fix $\xi$ and plot for different values of the ratio mentioned above. While plotting the curves we assume that $\Pi^{L,R}_1\ll\Pi^{L,R}_0$ and the mixing angle $\theta_{\rm mix}<0.25$ is respected.}
\label{nmchm_fig1}
\end{figure}
Due to the doublet-singlet mixing, the state corresponding to the observed Higgs field is
\begin{eqnarray}
\label{nmchm_2}
h^\prime&=&\cos\theta_{\rm{mix}}h_n-\sin\theta_{\rm{mix}}\eta_n ~,
\end{eqnarray}
where $\theta_{\rm mix}$ denotes the amount of mixing and is constrained by the LHC Higgs data. For the case, where both $m_\eta\gg m_{h}$ and $\langle \eta\rangle\gg\langle h\rangle$, the mixing angle can be simply parametrized as \cite{Sanz:2017tco}
\begin{equation}
\label{nmchm_3}
\theta_{\rm mix}\sim\frac{\langle h\rangle\langle\eta\rangle}{m_\eta^2}\ll 1~.
\end{equation}  
We also observe that
\begin{equation}
\label{nmchm_4}
k_{t\overline{t}\eta_n}\propto-\sqrt{\frac{\xi\chi}{1-\chi}}~,
\end{equation}
where $\chi=\langle\eta\rangle^2/f^2$. Because of an inherent $\mathcal{Z}_2$ symmetry associated with our choice of embedding, $\eta$ couples with the top quark as $\eta^2$. When the $\mathcal{Z}_2$ symmetry is spontaneously broken the dependence on $\chi$ appears. The appearance of $\xi$ is a consequence of constructing an $\rm SU(2)$ invariant Yukawa-like term involving the $\eta$ field. Finally the expression for the Yukawa coupling modifier involving the  observed Higgs is obtained as  
\begin{equation}
\label{nmchm_5}
k_{t\bar{t}h^\prime}=\cos\theta_{\mathrm{mix}}k_{t\bar{t}h_n}-\sin\theta_{\mathrm{ mix}}k_{t\bar{t}\eta_n}~.
\end{equation}
We show some representative plots illustrating the impact of $\chi$ on the Yukawa coupling modifier. In Fig.~\ref{nmchm_fig1} we present the variation of $k_{t\overline{t}h^\prime}$ with $\chi$ for $\rm NMCHM_{6_L-6_R}$. 
Obviously extra model dependence appears in the case of symmetric representation ($\bf 20$), where more than one Yukawa invariant exist.

% % % % % % % % % % % % % % % % % % % % % % % % % % % % % % % % % % % % % %

\section{Effective Phenomenological Lagrangian}
\label{eff_lag}

The modifications in the Higgs couplings as discussed in the previous section have two generic features: (\emph{i}) modification in $VVh$ coupling, arising from the non-linearity of the pNGBs, is universal as long as the coset belongs to $SO(N)/SO(N-1)$  group (modulo the mixing with other states), and (\emph{ii}) modification of the Yukawa couplings  depends on the choice of fermion embeddings. These can be captured in an effective Lagrangian as,
\begin{eqnarray}
\label{pheno_1}
\nonumber
\mathcal{L}&=&\mathcal{L}_{SMEFT}+\Delta\mathcal{L}~, \\
\mathcal{L}_{SMEFT}&\supset&\partial_\mu H^\dagger\partial^\mu H+\frac{g^2}{2}H^\dagger H\left(W_\mu W^\mu+\frac{1}{2\cos\theta_w^2}Z_\mu Z^\mu\right)-\sum_{u}y_u \overline{q}_LH^cu_R-\sum_{d}y_d \overline{q}_LHd_R  \nonumber \\
\label{pheno_2}
&-&\sum_{i=u,d}\frac{\alpha_s}{8\pi}y_{i}b^s_{i}\frac{H^\dagger H}{v^2}G_{\mu\nu}G^{\mu\nu}-\frac{\alpha_{em}}{8\pi}\left(\sum_{i=u,d}y_{i}b^{em}_{i}-g b^{em}_W\right)\frac{H^\dagger H}{v^2}F_{\mu\nu}F^{\mu\nu}. \nonumber \\
\end{eqnarray}  
Above, $\mathcal{L}_{SMEFT}$ comprises of the Standard Model effective Lagrangian with relevant dimension-4 and dimension-6 operators, where the explicit forms of the numerical coefficients $b_i$ are given in \cite{Ellis:2012rx}. In the $\rm SO(5)/SO(4)$ models, additional contributions to dimension-6 operators emerge, given by
\begin{eqnarray}
\nonumber
\Delta\mathcal{L}&\supset&\frac{1}{2f^2}\partial_\mu (H^\dagger H)\partial^\mu (H^\dagger H)-\sum_{u}(\Delta_u^\prime+\delta_u) y_u \frac{H^\dagger H}{f^2}\overline{q}_LH^cu_R-\sum_{d}(\Delta_d^\prime+\delta_d) y_d \frac{H^\dagger H}{f^2}\overline{q}_LHd_R\\
\label{pheno_3}
&-&\sum_{i=u,d}\frac{\alpha_s}{8\pi}\Delta_i^\prime y_{i}b^s_{i}\frac{H^\dagger H}{f^2}G_{\mu\nu}G^{\mu\nu}-\frac{\alpha_{em}}{8\pi}\sum_{i=u,d}\Delta_i^\prime y_{i}b^{em}_{i}\frac{H^\dagger H}{f^2}F_{\mu\nu}F^{\mu\nu}.
\end{eqnarray}
In the above Lagrangian we have dropped terms which are highly constrained by the electroweak precision observables \cite{Giudice:2007fh}. One can read off the Yukawa and $ggh/\gamma\gamma h$ coupling modifiers as 
\begin{equation}
\label{pheno_4}
k_{f\bar{f}h}=1+\left(\Delta^\prime_f+\delta_f-\frac{1}{2}\right) \xi\equiv 1+\left(\Delta_f+\delta_f\right) \xi, ~~~~~~~k^{(f)}_{ggh/\gamma\gamma h}=1+\left(\Delta^\prime_f-\frac{1}{2}\right)\xi\equiv 1+\Delta_f\xi~,
\end{equation}
and the modifier for $VVh$ coupling as
\begin{equation}
\label{pheno_8}
k_{VVh} = 1-\frac{1}{2}\xi~.
\end{equation}
While $\xi$ represents the ratio of the weak scale to the effective
scale of the theory, thus naturally controlling the coupling
modifiers, a brief discussion of the other two parameters, namely
$\Delta$ and $\delta$, in the effective Lagrangian is in order. The
origin of $\Delta$ can be traced back to the nonlinear realization of
the pNGB sector. In scenarios containing only one Yukawa invariant
this is a numerical constant (e.g. in $\rm MCHM_{5_L-5_R}$, $\Delta
\simeq - 3/2$), while in the extended models with several invariants
this factor may deviate depending on the details of the strong sector
resonances. In contrast, $\delta$ reflects the effect of partial
composite nature of the top quark in these theories, contributing to
the anomalous dimension of the top quark. In the effective $ggh$ and
$\gamma\gamma h$ vertex, in fact, contributions from the wave function
renormalization cancel against the resonance loop contributions
\cite{Azatov:2011qy, Montull:2013mla}. In our phenomenological
analysis, that follows in the next section, we will employ the
effective Lagrangian (Eqs.~\eqref{pheno_1} and \eqref{pheno_3}), to
confront the LHC Higgs data. All fitting are done assuming $\Delta_t$
and $\delta_t$ to be free parameters. Further we assume that the
bottom and $\tau$ Yukawa couplings are modified only by the universal
factor $\Delta_b=\Delta_\tau=-3/2$, \emph{i.e.} they are always
suppressed compared to their SM values. We also make a reasonable
approximation $\delta_b=\delta_\tau=0$. A complete list of all the
coupling modifiers within the $\rm SO(5)/SO(4)$ model is given in
Table~\ref{pheno_table1} \footnote{In
  \cite{Khachatryan:2016vau}, effective $ggh$ and
  $\gamma\gamma h$ coupling modifiers have been calculated keeping only the
  dominant terms:
  $k_{ggh} \simeq 1.06~(k_{ggh}^{(t)})^2 + 0.01~(k_{ggh}^{(b)})^2 - 0.07~
  k_{ggh}^{(b)} k_{ggh}^{(t)}$ and $k_{\gamma\gamma h} \simeq 0.07~
  (k_{\gamma\gamma h}^{(t)})^2 + 1.59~k_{WWh}^2- 0.66~k_{\gamma\gamma
    h}^{(t)} k_{WWh}$.}. The explicit expressions of $\Delta_t$ and
$\delta_t$ in terms of the form factors are defined in
Table~\ref{exp_model_table2} of Appendix \ref{delta}.
\begin{table}
\begin{longtable}{|c|c|}
\hline
 Modifiers & Dependence on parameters\\
\hline

\rule[-2ex]{0pt}{5.5ex} $k_{VVh}$ $(VV=WW, ZZ)$ & $1-\frac{1}{2}\xi$\\ 

\rule[-2ex]{0pt}{5.5ex} $k_{t\overline{t}h}$ & $1+\left(\Delta_t+\delta_t\right) \xi$\\ 

\rule[-2ex]{0pt}{5.5ex} $k_{ggh/\gamma\gamma h}^{(t)}$ & $1+\Delta_t \xi$\\ 

\rule[-2ex]{0pt}{5.5ex} $k_{b\overline{b}h}$ & $1-\frac{3}{2}\xi$\\ 

\rule[-2ex]{0pt}{5.5ex} $k_{ggh}^{(b)}$ & $1-\frac{3}{2}\xi$\\ 

\rule[-2ex]{0pt}{5.5ex} $k_{\tau\overline{\tau} h}$ & $ 1-\frac{3}{2}\xi$\\

%\rule[-2ex]{0pt}{5.5ex} $k_{ggh}$ & $1.058 (k_{ggh}^{(t)})^2+ 0.007 (k_{ggh}^{(b)})^2 - 0.065 k_{ggh}^{(b)} k_{ggh}^{(t)}$ \cite{Khachatryan:2016vau}\\ 

%\rule[-2ex]{0pt}{5.5ex} $k_{\gamma\gamma h}$ & $0.071 (k_{\gamma\gamma h}^{(t)})^2 + 1.589 k_{hWW}^2- 0.674 k_{\gamma\gamma h}^{(t)} k_{hWW}$\cite{Khachatryan:2016vau}\\ 

\hline 

\caption{\small\it Scaling of the Higgs effective couplings for $\rm SO(5)/SO(4)$ model.}
\label{pheno_table1}
\end{longtable}
\end{table}

The main feature that gets added when one moves to the next-to-minimal model is the presence of an additional singlet scalar and its mixing with the Higgs doublet. A description of the composite models including a SM singlet in the context of a strongly interacting light Higgs can be found in \cite{Chala:2017sjk}. Here we present a  simplified  effective Lagrangian keeping only the dominant terms.  We add the following piece involving $\eta$ to Eqs.~\eqref{pheno_2} and \eqref{pheno_3},
\begin{equation}
\label{pheno_5}
\Delta\mathcal{L}_{\eta}\sim \frac{1}{2}\partial_\mu\eta\partial^\mu\eta-\sum_{u} y_{u}(\Delta_{u}^{\eta})^\prime  \frac{\eta^2}{f^2}\overline{q}_LH^cu_R-\sum_{d} y_{d}(\Delta_{d}^{\eta})^\prime  \frac{\eta^2}{f^2}\overline{q}_LHd_R~.
\end{equation} 
Note that the dimension-5 operators involving a single $\eta$ field is not allowed in the presence of a $\mathcal{Z}_2$ symmetry, as discussed in previous section. Due to doublet-singlet scalar mixing ($\theta_{\rm mix}$), the Yukawa modifier for the observed Higgs boson ($h^\prime$) assumes the following form 
\begin{equation}
\label{pheno_6}
k_{f\bar{f}h^\prime}=\cos\theta_{\rm mix}\left(1+(\Delta_f+\delta_f)\xi\right)+\sin\theta_{\rm mix}\Delta^{\eta}_f\sqrt{\xi}~,
\end{equation}
where $\Delta^{\eta}_f$ is a function of $(\Delta^{\eta}_f)^\prime$ and the $\eta$-vev. The expressions for $\Delta^\eta_t$ for different representations are given in Table~\ref{nmchm_delta_table} of Appendix \ref{delta}. In the following analysis we assume $\Delta^{\eta}_t=\Delta^{\eta}_b$ for simplicity. 

The $VVh^\prime$ coupling modifier now picks up the additional factor $\cos\theta_{\rm mix}$ compared to the minimal coset (see Eq.~\eqref{pheno_8})
\begin{equation}
\label{pheno_7}
k_{VVh^\prime}=\cos\theta_{\rm mix}\sqrt{1-\xi}.
\end{equation}

% % % % % % % % % % % % % % % % % % % % % % % % % % % % % % % % % % % % % %

\section{Constraints from LHC Run 1 and Run 2 Higgs data}
\label{pheno}

In this section we discuss how the Higgs coupling modifications confront the recent LHC data \cite{Khachatryan:2016vau,ATLAS-CONF-2016-112,Aaboud:2017xsd,ATLAS-CONF-2017-043,ATLAS-CONF-2017-045,ATLAS-CONF-2017-076,ATLAS-CONF-2017-077,CMS-PAS-HIG-16-021,CMS-PAS-HIG-16-040,Sirunyan:2017exp,Sirunyan:2017khh,Sirunyan:2017elk,CMS-PAS-HIG-17-003,CMS-PAS-HIG-17-004}.  We perform a $\chi^2$ fit  to  assess the present constraints starting from the effective Lagrangian introduced in the previous section. We use the combined ATLAS+CMS Run 1 results for signal strengths, given by the `six-parameter' fit as shown in Table 15 of \cite{Khachatryan:2016vau}.  The so far available Run 2 (13 TeV) results are summarized in Table~\ref{data_table2}.

%\newpage
\begin{footnotesize}
%\begin{table}[t]
\begin{longtable}{|c|c|c|c|c|}

\hline 

\multicolumn{5}{|c|}{Run 2 Data}\\

\hline

\rule[-2ex]{0pt}{5.5ex} Collaboration & References & Decay Channels  & Production Modes & Results \\ 

\hline

\rule[-2ex]{0pt}{5.5ex}  & \cite{ATLAS-CONF-2016-112} & $WW$ & $VBF$ & $1.70^{+1.10}_{-0.90}$\\  
 
\cline{2-5} \rule[-2ex]{0pt}{5.5ex}
&  & & $ggF$ & $0.80^{+0.19}_{-0.18}$\\ 

\cline{4-5} \rule[-2ex]{0pt}{5.5ex}
& \cite{ATLAS-CONF-2017-045,ATLAS-CONF-2017-077} & $\gamma\gamma$ & $VBF$ & $2.10^{+0.60}_{-0
.60}$\\ 

\cline{4-5} \rule[-2ex]{0pt}{5.5ex}
&  & & $VH$ & $0.70^{+0.90}_{-0.80}$\\ 

\cline{4-5} \rule[-2ex]{0pt}{5.5ex}
& & & $t\overline{t}H$ & $0.60^{+0.70}_{-0.60}$\\ 

\cline{2-5} \rule[-2ex]{0pt}{5.5ex}
ATLAS & \cite{ATLAS-CONF-2017-043} & $ZZ^*$ & $ggF$ & $1.11^{+0.25}_{-0.22}$\\  

\cline{4-5} \rule[-2ex]{0pt}{5.5ex}
& & & $VBF$ & $4.00^{+1.77}_{-1.46}$\\  

\cline{2-5} \rule[-2ex]{0pt}{5.5ex}
& \cite{Aaboud:2017xsd,ATLAS-CONF-2017-077} & $b\overline{b}$ & $VH$ & $1.20^{+0.42}_{-0.36}$\\ 

\cline{4-5} \rule[-2ex]{0pt}{5.5ex}
& & & $t\overline{t}H$ & $0.80^{+0.60}_{-0.60}$\\ 
 
\cline{2-5} \rule[-2ex]{0pt}{5.5ex}
& \cite{ATLAS-CONF-2017-076,ATLAS-CONF-2017-077} & Multileptons & $t\overline{t}H$ & $1.60^{+0.50}_{-0.40}$\\ 

\hline 

\rule[-2ex]{0pt}{5.5ex}  & \cite{CMS-PAS-HIG-16-021} & $WW$ & $ggF+VBF+VH$  & $1.050^{+0.26}_{-0.26}$ \\ 

\cline{2-5} \rule[-2ex]{0pt}{5.5ex}
& & & $ggF$  & $1.11^{+0.19}_{-0.18}$\\ 

\cline{4-5} \rule[-2ex]{0pt}{5.5ex}
& \cite{CMS-PAS-HIG-16-040} & $\gamma\gamma$ & $VBF$ & $0.50^{+0.60}_{-0.50}$\\ 

\cline{4-5} \rule[-2ex]{0pt}{5.5ex}
&  & & $VH$ & $2.30^{+1.10}_{-1.00}$\\ 

\cline{4-5} \rule[-2ex]{0pt}{5.5ex}
&  & & $t\overline{t}H$ & $2.20^{+0.90}_{-0.80}$\\ 

\cline{2-5} \rule[-2ex]{0pt}{5.5ex}
CMS & \cite{Sirunyan:2017exp} & $ZZ^*$ & $ggF+t\overline{t}H$ & $1.20^{+0.35}_{-0.31}$\\  

\cline{2-5} \rule[-2ex]{0pt}{5.5ex}
& \cite{Sirunyan:2017khh} & $\tau\tau$ & $ggF+VBF+VH$ & $1.06^{+0.25}_{-0.24}$\\ 

\cline{2-5} \rule[-2ex]{0pt}{5.5ex}
& \cite{Sirunyan:2017elk} & $b\overline{b}$ & $ZH$ & $1.20^{+0.40}_{-0.40}$\\ 

\cline{2-5} \rule[-2ex]{0pt}{5.5ex}
& \cite{CMS-PAS-HIG-17-003} & $\tau_{h}$+others & $t\overline{t}H$ & $0.72^{+0.62}_{-0.53}$\\ 
\cline{2-5} \rule[-2ex]{0pt}{5.5ex}
& \cite{CMS-PAS-HIG-17-004}  & Multileptons & $t\overline{t}H$ & $1.50^{+0.50}_{-0.50}$\\ 
\hline

\caption{\small\it Results from the ATLAS and CMS collaborations for Higgs signal strengths at 13 TeV are tabulated. }
\label{data_table2}
\end{longtable}
%\end{table}
\end{footnotesize}
\newpage
The effective Lagrangian given in Eq.~\eqref{pheno_3}, which corresponds to $\rm SO(5)/SO(4)$ coset, have three independent  parameters $\xi$, $\Delta_t$ and $\delta_t$.   Using this  parametrization we calculate the Higgs signal strengths in various final states normalized to their SM values.  These are then  compared with the data using the $\chi^2$ fit. The minimum value of the $\chi^2$ and corresponding best-fit values of the parameters in the extended models are given below.
\begin{itemize}

\item Run 1 :
\begin{equation}
%\label{pheno_5}
\chi^2_{\rm min}=1.92, ~~\Delta_t=-0.31, ~~\delta_t=0.10, ~~\xi=0.13~,
\end{equation}

\item Run 1 + Run 2 :
\begin{equation}
%\label{pheno_6}
\chi^2_{\rm min}=18.85, ~~\Delta_t=-0.06, ~~\delta_t=0.02, ~~\xi=0.05~.
\end{equation}

\end{itemize}

This may be compared with a similar fit obtained for the  $\rm MCHM_{5_L-5_R}$ with only one  free parameter $\xi$. The best-fit values are given by
\begin{itemize}

\item Run 1 :
\begin{equation}
%\label{pheno_7}
\chi^2_{\rm min}=3.43, ~~\xi=0.007~,
\end{equation}

\item Run 1 + Run 2 :
\begin{equation}
%\label{pheno_8}
\chi^2_{\rm min}=19.72, ~~\xi=0.00~.
\end{equation}

\end{itemize}

\begin{figure}[]
\centering
\begin{subfigure}[t]{0.449\textwidth}
\includegraphics[width=\linewidth]{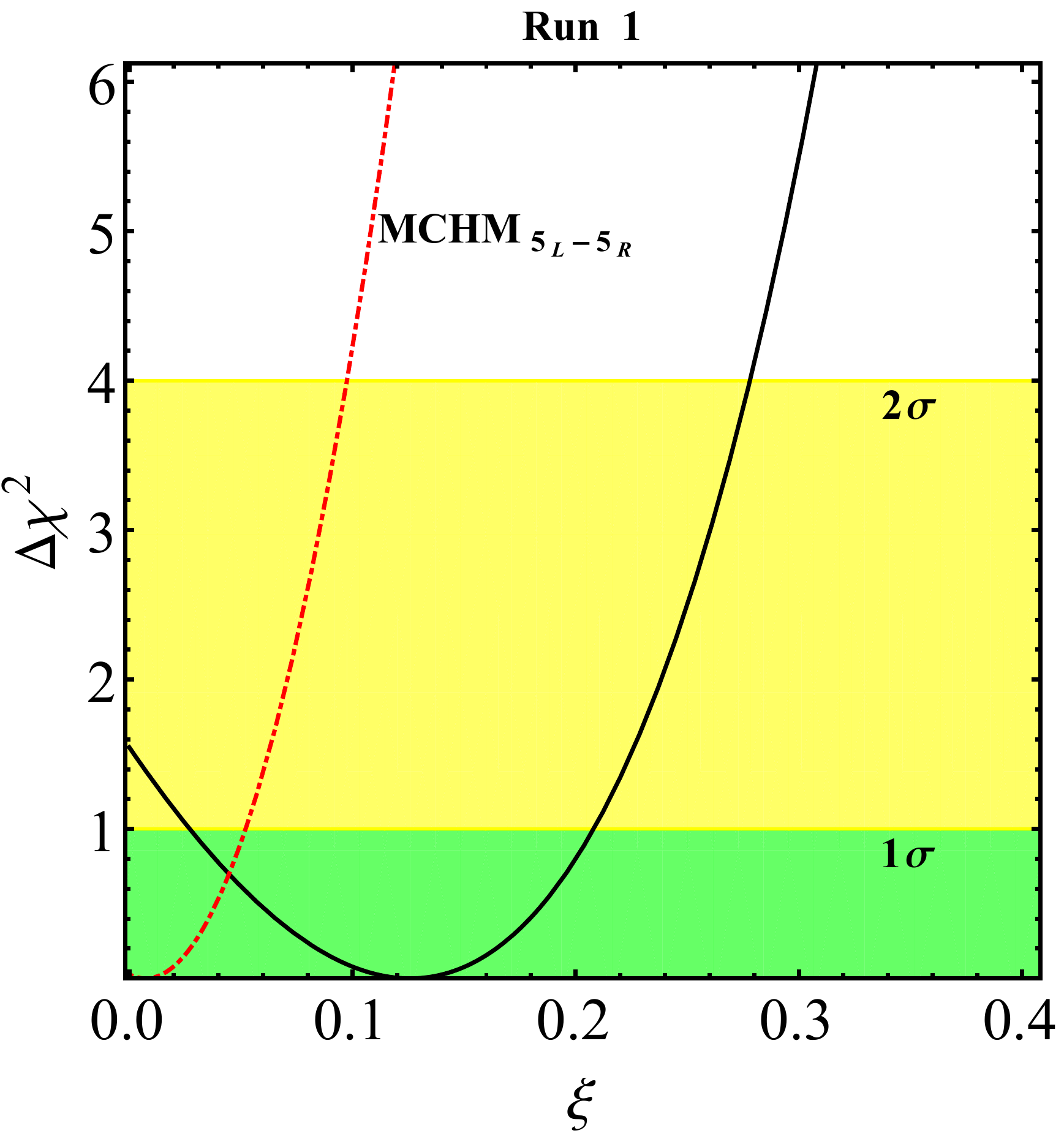}
%\caption{}
\end{subfigure} %
~
\begin{subfigure}[t]{0.45\textwidth}
\centering
\includegraphics[width=\linewidth]{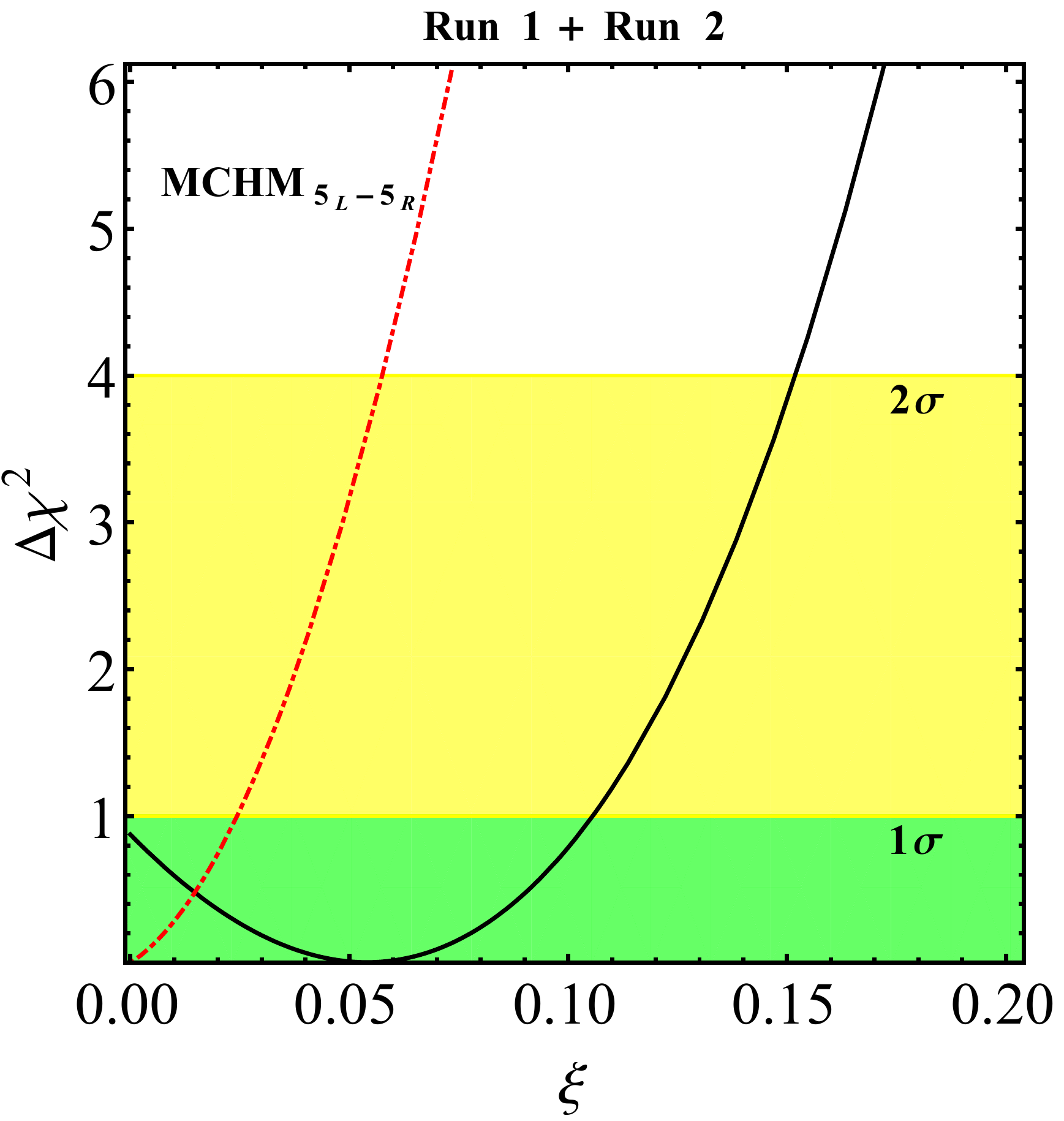}
%\caption{}
\end{subfigure} %
\caption{\small\it Results of $\chi^2$ analysis for Run 1 and Run 1 + Run 2 datasets are shown in the left and right panels, respectively. Solid black line represents $\Delta\chi^2=\chi^2-\chi^2_{\rm min}$ for the extended models, while the red dashed lines represents the same for $\rm MCHM_{5_L-5_R}$. Green and yellow regions denote the allowed range for $\xi$ at 68\% and 95\% CL, respectively.}
\label{pheno_fig1}
\end{figure}

In Figs.~\ref{pheno_fig1} we plot $\Delta\chi^2$ as a function of
$\xi$, corresponding to our effective Lagrangian Eq.~\eqref{pheno_3},
with all other parameters fixed to their best-fit values.  For
comparison, we also show the curve for $\rm MCHM_{5_L-5_R}$, and our
results agree with \cite{Sanz:2017tco} wherever we overlap. For the
extended models, we obtain a lower bound $f\geq 465~\rm GeV$ at 95\%
CL from Run 1 data only. This should be compared with $f\geq 780$ GeV
at 95\% CL for $\rm MCHM_{5_L-5_R}$. The relaxation of the bounds on
$f$ for the extended models follows from the reduced correlation
between $k_{t\overline{t}h}$ and $k_{VVh}$ in the effective Lagrangian
in Eq.~\eqref{pheno_3} as compared to the tight correlation in $\rm
MCHM_{5_L-5_R}$. We find that combined Run 1 and Run 2 data give
significantly more stringent lower bound, namely, $f\geq 640$ GeV at
95\% CL for the extended models.

\begin{figure}[]
\centering
\begin{subfigure}[t]{0.45\textwidth}
\includegraphics[width=\linewidth]{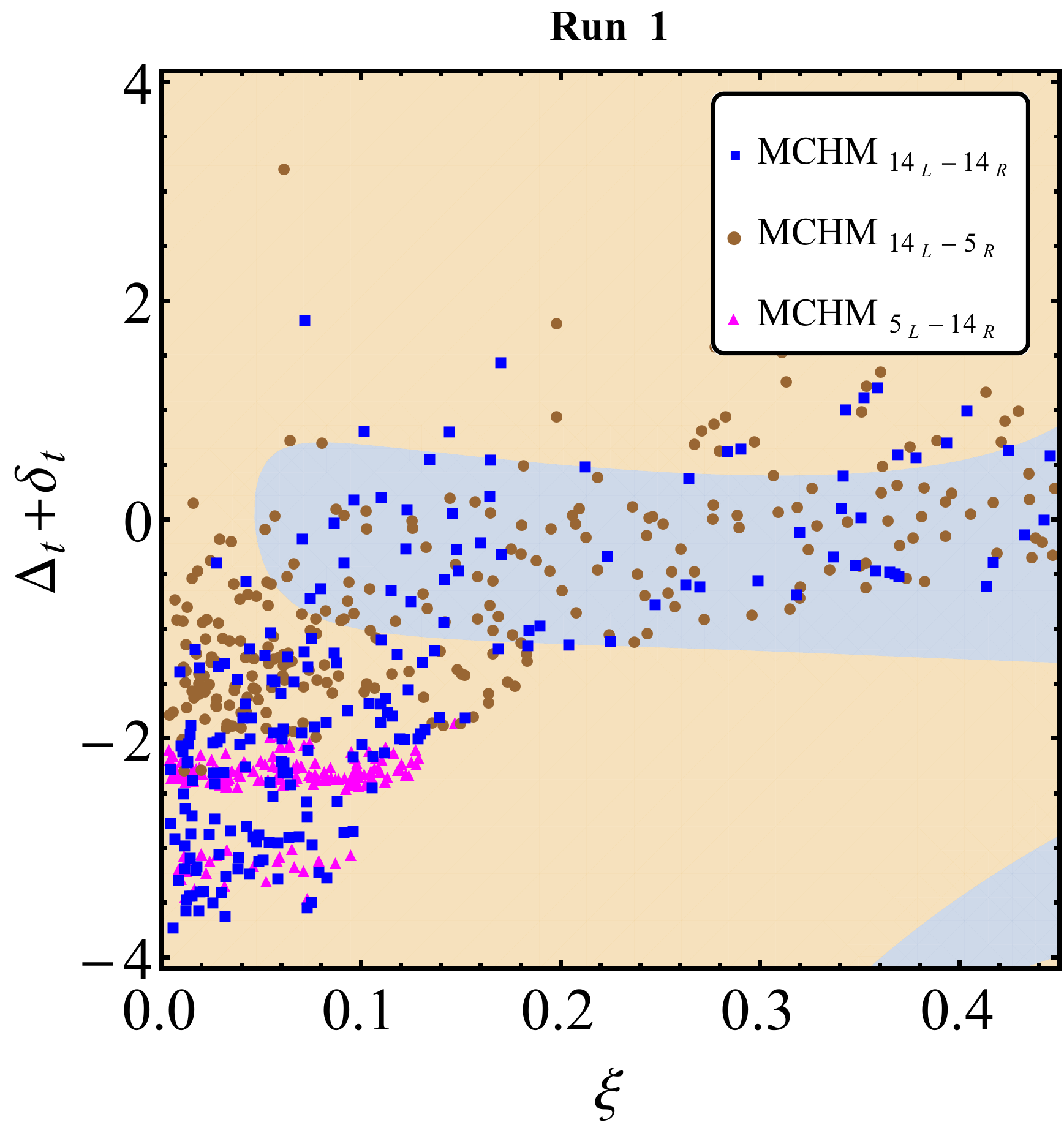}
%\caption{}
\end{subfigure} %
~
\begin{subfigure}[t]{0.45\textwidth}
\centering
\includegraphics[width=\linewidth]{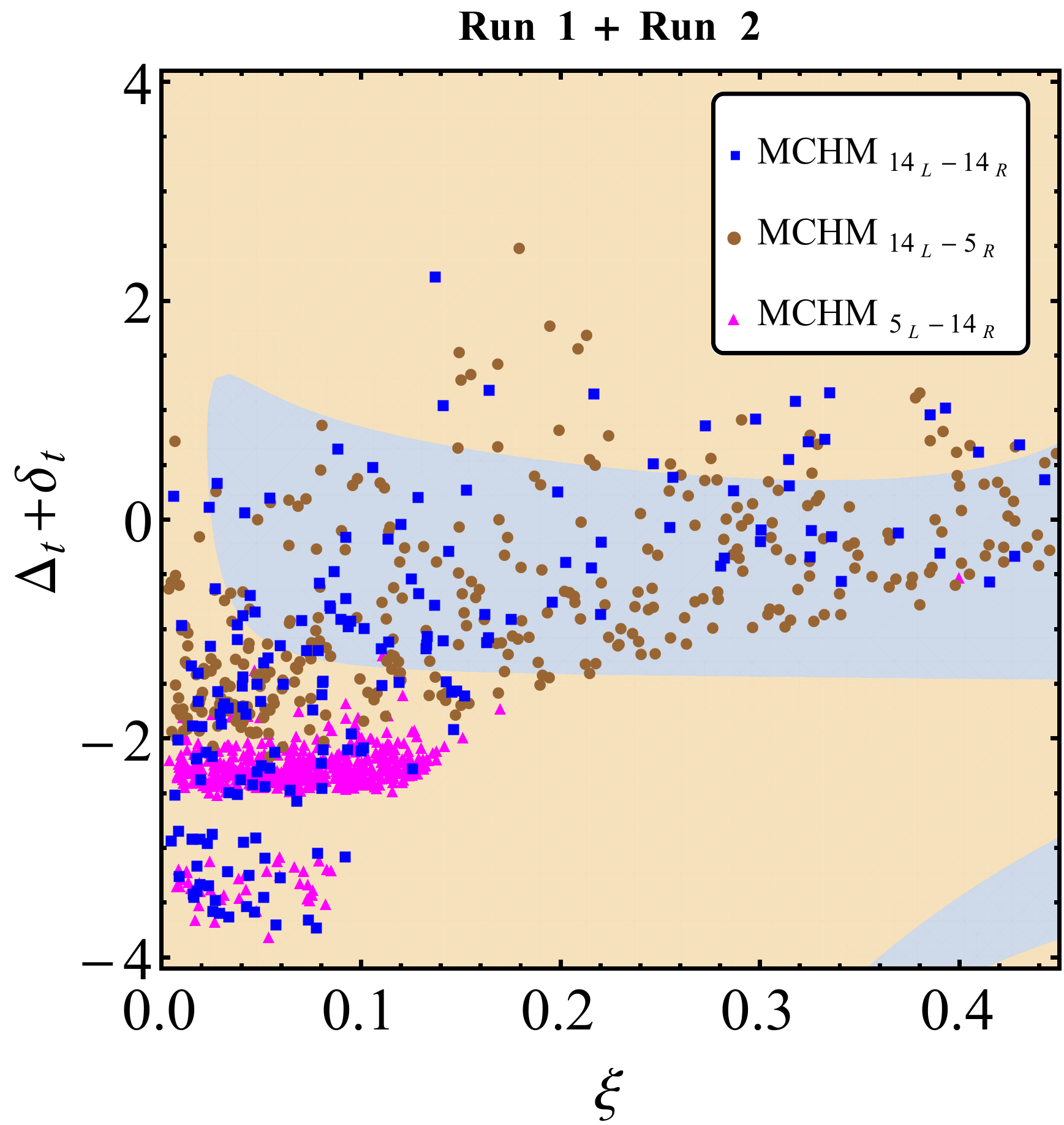}
%\caption{}
\end{subfigure} %
\caption{ \small\it In both the left and right panels, blue shaded
  regions denote relatively lower values of $\chi^2/d.o.f.$ for the
  extended models, given by the effective Lagrangian in
  Eq.~\eqref{pheno_3}, compared to the $\rm MCHM_{5_L-5_R}$.  Model
  points, with resonance masses and decay constants as inputs,
  satisfying the constraints of Eq.~\eqref{exp_model_9}, are
  superimposed.}
\label{pheno_fig2}
\end{figure}

\begin{figure}[]
\centering
\begin{subfigure}[t]{0.44\textwidth}
\includegraphics[width=\linewidth]{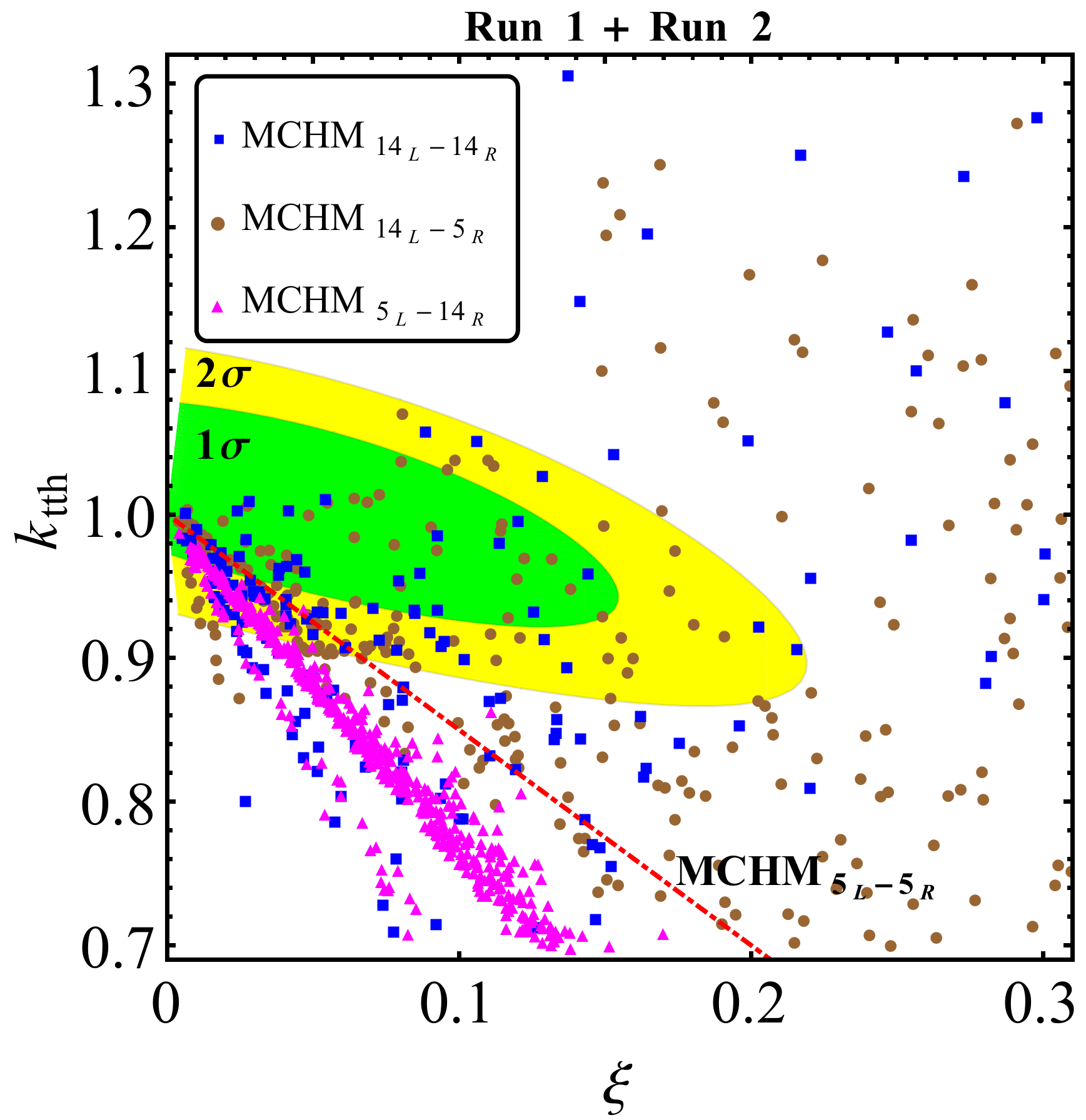}
%\caption{}
\end{subfigure} %
~
\begin{subfigure}[t]{0.45\textwidth}
\centering
\includegraphics[width=\linewidth]{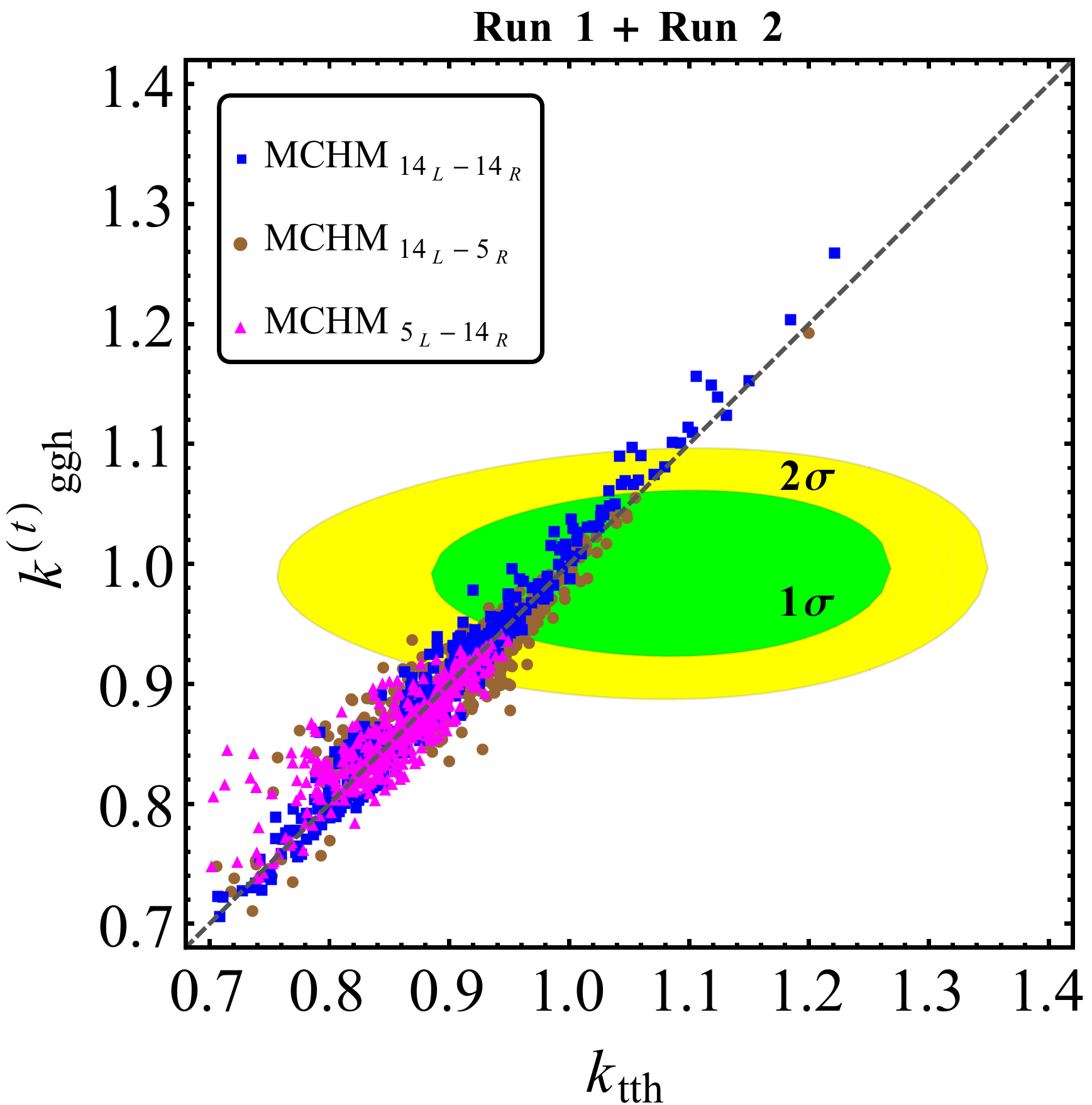}
%\caption{}
\end{subfigure} %
\caption{\small\it We present the regions in the
  $k_{t\overline{t}h}-\xi$ and $k_{ggh}^{(t)}-k_{t\overline{t}h}$
  planes allowed at 68\% (green) and 95\% (yellow) CL using combined
  Run 1 and available Run 2 data.  In the left panel the red line
  corresponds to $\rm MCHM_{5_L-5_R}$. On the right panel, the grey
  dashed line corresponds to $\delta_t = 0$. Valid `extended model'
  points are observed to lie within the experimentally allowed
  regions.}
\label{pheno_fig3}
\end{figure}

In Figs.~\ref{pheno_fig2} we check whether the extended models fit the
data better ({\em i.e.} $\chi^2/d.o.f.$ is lower) than $\rm
MCHM_{5_L-5_R}$.  In the blue shaded region the extended models, as
parametrically encoded in the effective Lagrangian in
Eq.~\eqref{pheno_3}, fit relatively better for the entire range of
$\xi$. On the same plot we also throw the actual model points, with
the resonance masses and decay constants as the strong sector inputs,
discussed on a case-by-case basis in Section \ref{minimal}, satisfying
the constraints shown in Eq.~\eqref{exp_model_9}.  In
Figs.~\ref{pheno_fig3}, the experimentally preferred regions for the
coupling modifiers are shown at 68\% and 95\% CL in the
($k_{t\overline{t}h}$--$\xi$) and ($k_{ggh}^{(t)}$--$k_{t\overline{t}h}$)
planes. The model points are observed to span over a large range of
the preferred regions. It may be noted that present experimental
precision is not sensitive to the value of $\delta_t$ separately; what
is in fact bounded is the combination $(\Delta_t+\delta_t)$. Future
colliders may have sufficient precision to sense the different
modifications in the top Yukawa coupling and the effective $ggh$
coupling.

\begin{figure}[]
\centering
\includegraphics[width=0.45\textwidth]{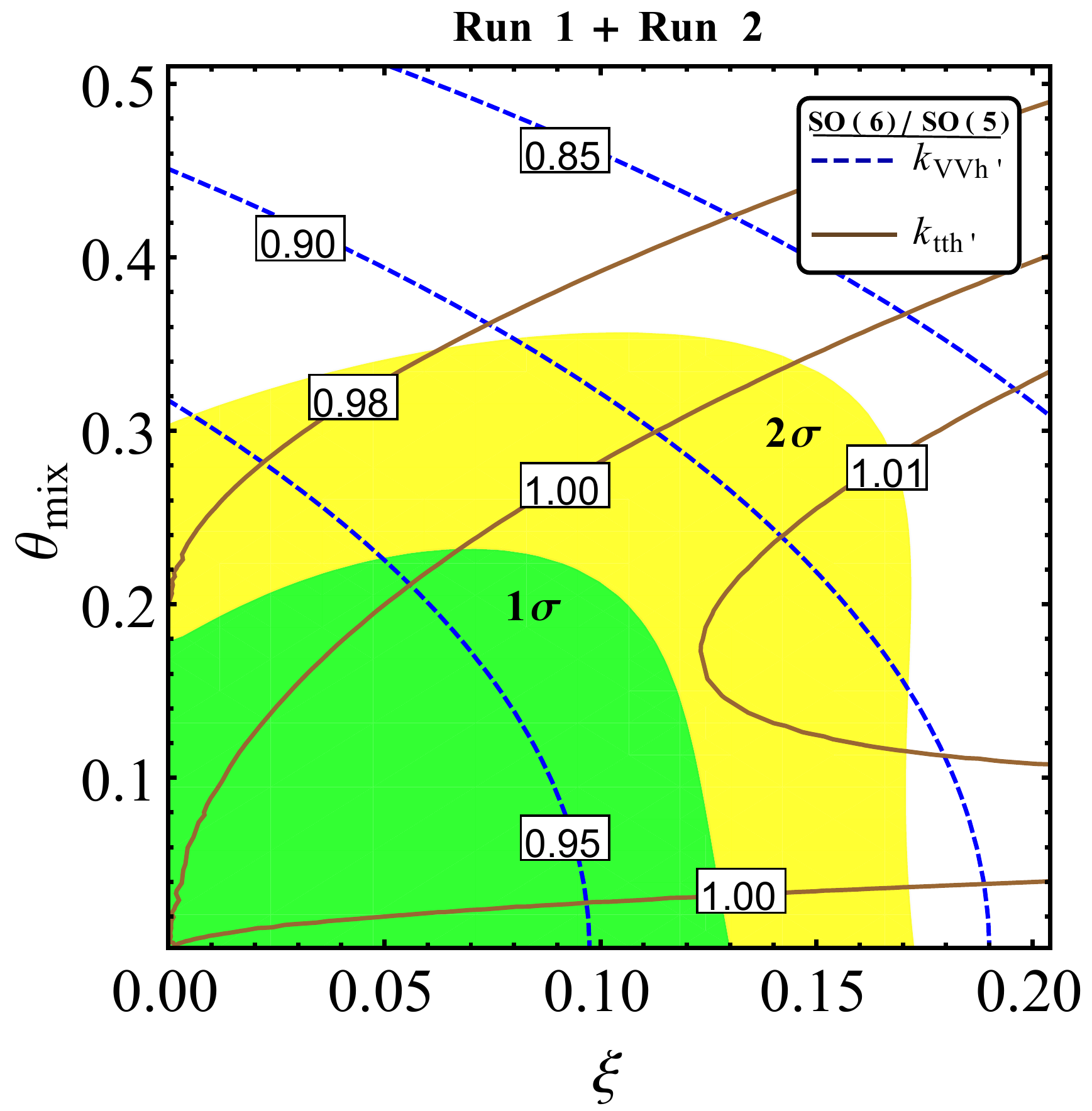}
\caption{\small\it For the next-to-minimal model, the allowed regions
  in the $\theta_{\mathrm{mix}}-\xi$ plane at 68\% (green) and 95\%
  (yellow) CL are shown using the combined Run 1 and available Run 2
  data. The brown lines represent the contours of fixed
  $k_{t\overline{t}h^\prime}$, while the blue dashed lines correspond
  to that of $k_{VVh^\prime}$.}
\label{pheno_fig4}
\end{figure}

Moving to the next-to-minimal coset, we deal with a new feature that
the Higgs doublet now mixes with a real singlet.  The mixing results
in a further suppression of the observed Higgs boson coupling to the
massive gauge bosons. The Yukawa couplings are modified too because of
the presence of a singlet.  We perform a similar $\chi^2$ analysis
with the combined Run 1 and Run 2 results to impose an upper bound on
the amount of mixing. Fig.~\ref{pheno_fig4} shows that the maximum
amount of mixing allowed so far at 95\% CL is $\theta_{\rm mix}\sim
0.35$. Future data would constrain it even further
\cite{Banerjee:2017qod}.

% % % % % % % % % % % % % % % % % % % % % % % % % % % % % % % % % % % % % %

\section{Conclusions}
\label{conc}

Non-linearity of pNGB dynamics modifies the Higgs boson couplings with
the weak gauge bosons as well as with the fermions compared to their
SM expectations.  The ratio $\xi$, which parametrizes the hierarchy
between the weak scale and the strong sector spontaneous symmetry
breaking scale, controls this deformation. In $\rm MCHM_{5_L-5_R}$,
the Yukawa sector contains a single invariant. Here, the single
parameter $\xi$ appears in the modifications of both $VVh$ and $f\bar
fh$ couplings, leading to a rather strong lower limit $f \geq 1$ TeV,
as the data show increasing affinity towards the SM predictions.  In
the extended models, $\rm MCHM_{14_L-14_R}$, $\rm MCHM_{14_L-5_R}$ and
$\rm MCHM_{5_L-14_R}$, owing to the presence of more than one
invariant in the Yukawa sector, the $f \bar fh$ coupling modifier
depends on other parameters of the strong sector in addition to $\xi$.
This releases the tension leading to a new lower limit $f \geq 640$
GeV, which is much relaxed compared to the limit in $\rm
MCHM_{5_L-5_R}$.

An important feature of these models is the emergence of a
parametric difference in the top Yukawa and the effective
gluon-gluon-Higgs vertices. This arises because of a cancellation
between the top-partner resonance masses in the loop with the wave
function renormalization of the top quark in the calculation of the
effective $ggh$ vertex. However, the present data is insensitive to
smell this difference.

We have in fact constructed a phenomenological Lagrangian which
captures the effects of a vast array of such models with different
fermionic representations. We have constrained the parameters of this
Lagrangian using LHC data and observed that the allowed regions are
quite consistent with a reasonable choice of strong sector input
parameters of the individual models which yield correct values of
$m_t$, $v$ and $m_h$.

We have extended our analysis to the next-to-minimal model as
well. The appearance of a real singlet scalar adds a new twist to
phenomenology, whose mixing with the Higgs doublet is constrained by
the LHC data. Interestingly, the singlet scalar also contributes to
the top Yukawa through an effective higher dimensional operator.

Our analysis shows that further precision, likely to be achieved in
future colliders, would constrain these scenarios to the extent that
individual models could be discriminated, and the proposition that the
Higgs boson may have a spatial extension would be challenged with more
ammunition.

%\acknowledgments
\begin{small}
\section*{Acknowledgments}
We thank S.~Bhattacharya, M.~R.~Gangopadhyay, S.~Roy Chowdhury,
K.~Mondal, C.~Mondal and D.~Das for discussions. AB acknowledges
financial support from Department of Atomic Energy, Government of
India. GB acknowledges support of the J.C. Bose National Fellowship
from the Department of Science and Technology, Government of India
(SERB Grant No. SB/S2/JCB-062/2016). Support from the Indo French
Center for Promotion of Advanced Research (CEFIPRA Project No. 5404-2)
is also acknowledged by GB and NK. TSR acknowledges the hospitality
provided by ICTP, Italy, under the Associate program, during the
initial stages of the project. TSR is partially supported by the
Department of Science and Technology, Government of India, under the
Grant Agreement number IFA13-PH-74 (INSPIRE Faculty Award).
\end{small}

% % % % % % % % % % % % % % % % % % % % % % % % % % % % % % % % % % % % % %

% % % % % % % % % % % % % % Appendix % % % % % % % % % % % % % % % % % % % %
\appendix
%\small
\section{Fermion Embeddings}
\label{embeddings}

% % % % % % % % % % % % % % % % % % % % % % % % % % % % % % % % % % % % % %

\subsection{$\rm\bf SO(5)/SO(4)$ Coset}
\label{mchm_embeddings}

Fundamental $\bf 5$ and symmetric $\bf 14$ representations of $\rm SO(5)$ can be decomposed under the unbroken $\rm SO(4)\equiv SU(2)_L\times SU(2)_R$ as follows:
\begin{eqnarray}
\label{embeddings_1}
\nonumber
\bf 5 &=& \bf 1\oplus 4 =(1,1)\oplus (2,2),\\
\bf 14 &=& \bf 1\oplus 4\oplus 9 =(1,1)\oplus (2,2) \oplus (3,3).
\end{eqnarray}
We embed $t_L$ into the $(\bf{2},\bf{2})$'s so that the correction to $Zb\overline{b}$ vertex is under control, while $t_R$ is embedded into the $(\bf{1},\bf{1})$.  
The embeddings of the top quarks into incomplete multiplets of $\bf 5$ and $\bf 14$ are given below. 
\begin{equation}
\label{embeddings_2}
Q_{t_L}^{5}=\left(\Psi_{(2,2)},0\right)^T,~~~
T_{t_R}^{5}=\left(0,0,0,0,t_R\right)^T, 
\end{equation} 
and
\begin{equation}
\label{embeddings_3}
Q_{t_L}^{14}=\left(
\begin{array}{c|c}
0_{4\times 4} &  \frac{\Psi_{(2,2)}^T}{\sqrt{2}} \\
\hline
\frac{\Psi_{(2,2)}}{\sqrt{2}} & 0
\end{array}
\right),~~~ 
T_{t_R}^{14}=
\left(
\begin{array}{c|c}
-\frac{t_R}{2\sqrt{5}}\mathnormal{I}_4 & 0_{4\times 1} \\
\hline
 0_{1\times 4} & \begin{array}{c}
4\frac{t_R}{2\sqrt{5}} 
\end{array}
\end{array}
\right),
\end{equation}
where
\begin{equation}
\label{embeddings_4}
\Psi_{(2,2)}=\frac{1}{\sqrt{2}}(ib_L, b_L, it_L, -t_L).
\end{equation}

% % % % % % % % % % % % % % % % % % % % % % % % % % % % % % % % % % % % %

\subsection{$\rm\bf SO(6)/SO(5)$ Coset}
\label{nmchm_embeddings}

Decomposition of different representations of $SO(6)$, used in the
main text, under the maximal subgroup $SO(6)\supset SO(4)\times
SO(2)\simeq SU(2)_L\times SU(2)_R\times U(1)_\eta$, is as follows:
\begin{eqnarray}
\label{embeddings_5}
\nonumber
\mathbf{6}_0 &=&(\mathbf{2},\mathbf{2})_0\oplus (\mathbf{1},\mathbf{1})_2\oplus (\mathbf{1},\mathbf{1})_{-2} ~,\\
\mathbf{15}_0 &=& (\mathbf{1},\mathbf{1})_0\oplus (\mathbf{2},\mathbf{2})_2\oplus (\mathbf{2},\mathbf{2})_{-2}\oplus (\mathbf{3},\mathbf{1})_0\oplus (\mathbf{1},\mathbf{3})_{0} ~,\\
\nonumber
\mathbf{20}_0 &=& (\mathbf{1},\mathbf{1})_0\oplus(\mathbf{1},\mathbf{1})_4\oplus(\mathbf{1},\mathbf{1})_{-4}\oplus (\mathbf{2},\mathbf{2})_2\oplus (\mathbf{2},\mathbf{2})_{-2}\oplus (\mathbf{3},\mathbf{3})_0~,
\end{eqnarray}
where the subscripts denote the charges under $\rm U(1)_\eta$. Embedding of $t_L$ and $t_R$ in the above representations are given as
\begin{equation}
\label{embeddings_6}
Q_{t_L}^{6}=\left(\Psi_{(2,2)}, 0, 0\right)^T~,~~~~~~
T_{t_R}^{6}=\left(0,0,0,0,0, t_R\right)^T,
\end{equation}
\begin{equation}
\label{embeddings_7}
Q_{t_L}^{15}=
\left(
\begin{array}{c|c}
0_{4\times 4} & \begin{array}{cc}
0 & \frac{\Psi_{(2,2)}^T}{\sqrt{2}}
\end{array}  \\
\hline
\begin{array}{rr}
0~~~~ \\ -\frac{\Psi_{(2,2)}}{\sqrt{2}}
\end{array} & 0_{2\times 2}
\end{array}
\right)~,~~~
T_{t_R}^{15}=
\left(
\begin{array}{c|c}
\begin{array}{c|c}
 \begin{array}{cc}
0 & -i\frac{t_R}{2}\\
i\frac{t_R}{2} & 0
\end{array} & 0_{2\times 2}  \\
\hline
0_{2\times 2} & \begin{array}{cc}
0 & i\frac{t_R}{2}\\
-i\frac{t_R}{2} & 0
\end{array}
\end{array} & 0_{4\times 2}
\\
\hline
 0_{2\times 4} & 0_{2\times 2}
\end{array}
\right),
\end{equation}
and 
\begin{equation}
\label{embeddings_8}
Q_{t_L}^{20}=\left(
\begin{array}{c|c}
0_{4\times 4} & \begin{array}{cc}
0 & \frac{\Psi_{(2,2)}^T}{\sqrt{2}}
\end{array}  \\
\hline
\begin{array}{rr}
0~~~ \\ \frac{\Psi_{(2,2)}}{\sqrt{2}}
\end{array} & 0_{2\times 2}
\end{array}
\right)~,~~~~~
T_{t_R}^{20}=
\left(
\begin{array}{c|c}
-\frac{t_R}{2\sqrt{3}}\mathnormal{I}_4 & 0_{4\times 2} \\
\hline
 0_{2\times 4} & \frac{t_R}{\sqrt{3}}\mathnormal{I}_2
\end{array}
\right).
\end{equation}

% % % % % % % % % % % % % % % % % % % % % % % % % % % % % % % % % % % % % %

\section{Details of Form Factors}
\label{form factors}

The form factors appearing in Table~\ref{exp_model_table1} can be
decomposed under unbroken $\rm SO(4)$ and written in terms of masses
and decay constants of the resonances.  Detailed expressions of the form factors for $\rm MCHM_{14_L-14_R}$, $\rm MCHM_{14_L-5_R}$ and $\rm MCHM_{5_L-14_R}$ models are listed below (for calculations, see \cite{Montull:2013mla}).

% % % % % % % % % % % % % % % % % % % % % % % % % % % % % % % % % % % % %  

\subsection*{$\rm\bf MCHM_{14_L-14_R}$}
\label{mchm14}

\begin{eqnarray}
%\label{mchm_14_2}
\begin{rcases}
\Pi^L_0&=1+\frac{|F^{L}_4|^2}{q^2+m_4^2},\\
\Pi^L_1&= \left(\frac{5}{4}\frac{|F^{L}_1|^2}{q^2+m_1^2}-\frac{5}{2}\frac{|F^{L}_4|^2}{q^2+m_4^2}+\frac{5}{4}\frac{|F^{L}_9|^2}{q^2+m_9^2}\right),\\
\Pi^L_2&= \left(-\frac{5}{4}\frac{|F^{L}_1|^2}{q^2+m_1^2}+2\frac{|F^{L}_4|^2}{q^2+m_4^2}-\frac{3}{4}\frac{|F^{L}_9|^2}{q^2+m_9^2}\right),\\
\Pi^R_0&= 1+\frac{|F^{R}_1|^2}{q^2+m_1^2},\\
\Pi^R_1&= \left(\frac{5}{2}\frac{|F^{R}_4|^2}{q^2+m_4^2}-\frac{5}{2}\frac{|F^{R}_1|^2}{q^2+m_1^2}\right),\\
\Pi^R_2&= \left(\frac{25}{16}\frac{|F^{R}_1|^2}{q^2+m_1^2}-\frac{5}{2}\frac{|F^{R}_4|^2}{q^2+m_4^2}+\frac{15}{16}\frac{|F^{R}_9|^2}{q^2+m_9^2}\right),\\
\Pi^{LR}_1&= -\frac{\sqrt{5}}{2}\left(\frac{F^{L}_1F^{R*}_1m_1}{q^2+m_1^2}-\frac{F^{L}_4F^{R*}_4m_4}{q^2+m_4^2}\right),\\
\Pi^{LR}_2&=- \left(-\frac{5\sqrt{5}}{8}\frac{F^{L}_1F^{R*}_1m_1}{q^2+m_1^2}+\sqrt{5}\frac{F^{L}_4F^{R*}_4m_4}{q^2+m_4^2}-\frac{3\sqrt{5}}{8}\frac{F^{L}_9F^{R*}_9m_9}{q^2+m_9^2}\right).
\end{rcases}
\end{eqnarray} 

% % % % % % % % % % % % % % % % % % % % % % % % % % % % % % % % % % % % % %

\subsection*{$\rm\bf MCHM_{14_L-5_R}$}
\label{mchm_14_5}

\begin{eqnarray}
%\label{mchm_14_5_2}
\begin{rcases}
\Pi^L_0&= 1+\frac{|F^L_4|^2}{q^2+m_4^2},\\
\Pi^L_1&= \frac{5}{4}\frac{|F^L_1|^2}{q^2+m_1^2}-\frac{5}{2}\frac{|F^L_4|^2}{q^2+m_4^2}+\frac{5}{4}\frac{|F^L_9|^2}{q^2+m_4^2},\\
\Pi^L_2&= 2\frac{|F^L_4|^2}{q^2+m_4^2}-\frac{5}{4}\frac{|F^L_1|^2}{q^2+m_1^2}-\frac{3}{4}\frac{|F^L_9|^2}{q^2+m_4^2},\\
\Pi^R_0&= 1+\frac{|F^R_1|^2}{q^2+m_1^2},\\
\Pi^R_1&= \frac{|F^R_4|^2}{q^2+m_4^2}-\frac{|F^R_1|^2}{q^2+m_1^2},\\
\Pi^{LR}_1&= \frac{1}{\sqrt{2}}\frac{F^L_4F^{R*}_4m_4}{q^2+m_4^2}-\frac{\sqrt{5}}{2}\frac{F^L_1F^{R*}_1m_1}{q^2+m_1^2},\\
\Pi^{LR}_2&= \frac{\sqrt{5}}{2}\frac{F^L_1F^{R*}_1m_1}{q^2+m_1^2}-\sqrt{2}\frac{F^L_4F^{R*}_4m_4}{q^2+m_4^2}.
\end{rcases}
\end{eqnarray}

% % % % % % % % % % % % % % % % % % % % % % % % % % % % % % % % % % % % % %

\subsection*{$\rm\bf MCHM_{5_L-14_R}$}
\label{mchm_5_14}

\begin{eqnarray}
%\label{mchm_5_14_1}
\begin{rcases}
\Pi^L_0&= 1+\frac{|F^L_4|^2}{q^2+m_4^2}\\
\Pi^L_1&= \frac{1}{2}\left(\frac{|F^L_1|^2}{q^2+m_1^2}-\frac{|F^L_4|^2}{q^2+m_4^2}\right)\\
\Pi^R_0&= 1+\frac{|F^R_1|^2}{q^2+m_1^2}\\
\Pi^R_1&= \frac{5}{2}\left(\frac{|F^R_4|^2}{q^2+m_4^2}-\frac{|F^R_1|^2}{q^2+m_1^2}\right)\\
\Pi^R_2&= \frac{25}{16}\frac{|F^R_1|^2}{q^2+m_1^2}-\frac{5}{2}\frac{|F^R_4|^2}{q^2+m_4^2}+\frac{15}{16}\frac{|F^R_9|^2}{q^2+m_4^2}\\
\Pi^{LR}_1&= \frac{\sqrt{5}}{2}\frac{F^L_4F^{R*}_4m_4}{q^2+m_4^2}-\frac{1}{\sqrt{2}}\frac{F^L_1F^{R*}_1m_1}{q^2+m_1^2}\\
\Pi^{LR}_2&= \frac{5\sqrt{2}}{8}\frac{F^L_1F^{R*}_1m_1}{q^2+m_1^2}-\frac{\sqrt{5}}{2}\frac{F^L_4F^{R*}_4m_4}{q^2+m_4^2}
\end{rcases}
\end{eqnarray}

% % % % % % % % % % % % % % % % % % % % % % % % % % % % % % % % % % % %

\section{$\Pi$-functions for Next-to-Minimal Models}
\label{nmchm_struc}

The $\Pi$-functions for the next-to-minimal case are given for
different representations in Table~\ref{nmchm_table1}. The $\rm
NMCHM_{15_L-1_R}$ case is not included in the table because it cannot
generate a Yukawa term in the Lagrangian.
%\begin{small}
%\begin{table}
%\begin{longtable}{|c|c|c|c|}
\begin{longtabu} to \linewidth {|c|P{3cm}|P{3.2cm}|P{5.2cm}|}
\hline

\rule[-2ex]{0pt}{5.5ex} 
Models & $\Pi_{t_L}(h,\eta)$ & $\Pi_{t_R}(h,\eta)$ & $\Pi_{t_Lt_R}(h,\eta)$  \\ 

\hline 

\rule[-2ex]{0pt}{5.0ex} 
$\rm NMCHM_{6_L-1_R}$ & $\Pi^L_0+\Pi^L_1 \frac{h^2}{f^2}$ & $\Pi^R_0$  & $\Pi^{LR}_1\frac{h}{f}$ \\ 

\hline 

\rule[-2ex]{0pt}{5.5ex} 
$\rm NMCHM_{6_L-6_R}$ & $\Pi^L_0+\Pi^L_1 \frac{h^2}{f^2}$ & $\Pi^R_0+\Pi^R_1 \frac{h^2}{f^2}+\Pi^R_\eta \frac{\eta^2}{f^2}$  & $\Pi^{LR}_1\frac{h}{f}\sqrt{1-\frac{h^2}{f^2}-\frac{\eta^2}{f^2}}$ \\ 

\hline 

\rule[-2ex]{0pt}{5.5ex} 
$\rm NMCHM_{6_L-15_R}$ & $\Pi^L_0+\Pi^L_1 \frac{h^2}{f^2}$ & $\Pi^R_0+\Pi^R_1 \frac{h^2}{f^2}$  & $\Pi^{LR}_1\frac{h}{f}$ \\ 

\hline 

\rule[-2ex]{0pt}{5.5ex} 
$\rm NMCHM_{6_L-20_R}$  & $\Pi^L_0+\Pi^L_1 \frac{h^2}{f^2}$ & $\Pi^R_0+\Pi^R_1 \frac{h^2}{f^2}+\Pi^R_2 \frac{h^4}{f^4}$ & $\frac{h}{f}\left(\Pi^{LR}_1+\Pi^{LR}_2 \frac{h^2}{f^2}\right)$ \\ 

\hline 

\rule[-2ex]{0pt}{5.5ex} 
$\rm NMCHM_{15_L-6_R}$ & $\Pi^L_0+\Pi^L_1 \frac{h^2}{f^2}+\Pi^L_\eta \frac{\eta^2}{f^2}$ &$\Pi^R_0+\Pi^R_1 \frac{h^2}{f^2}+\Pi^R_\eta \frac{\eta^2}{f^2}$  & $\Pi^{LR}_1\frac{h}{f}$ \\ 

\hline 

\rule[-2ex]{0pt}{5.5ex} 
$\rm NMCHM_{15_L-15_R}$ & $\Pi^L_0+\Pi^L_1 \frac{h^2}{f^2}+\Pi^L_\eta \frac{\eta^2}{f^2}$ & $\Pi^R_0+\Pi^R_1 \frac{h^2}{f^2}$  & $\Pi^{LR}_1\frac{h}{f}\sqrt{1-\frac{h^2}{f^2}-\frac{\eta^2}{f^2}}$ \\ 

\hline 

\rule[-2ex]{0pt}{5.5ex} 
$\rm NMCHM_{15_L-20_R}$ & $\Pi^L_0+\Pi^L_1 \frac{h^2}{f^2}+\Pi^L_\eta \frac{\eta^2}{f^2}$ & $\Pi^R_0+\Pi^R_1 \frac{h^2}{f^2}+\Pi^R_2 \frac{h^4}{f^4}$  & $\Pi^{LR}_1\frac{h}{f}\sqrt{1-\frac{h^2}{f^2}-\frac{\eta^2}{f^2}}$ \\ 

\hline 

\rule[-2ex]{0pt}{5.5ex} 
\multirow{2}{*}{$\rm NMCHM_{20_L-1_R}$} & $\Pi^L_0+\Pi^L_1 \frac{h^2}{f^2}+\Pi^L_2 \frac{h^4}{f^4}$ & \multirow{2}{*}{$\Pi^R_0$}  & \multirow{2}{*}{$\Pi^{LR}_1\frac{h}{f}\sqrt{1-\frac{h^2}{f^2}-\frac{\eta^2}{f^2}}$} \\
& $+\Pi^L_\eta \frac{\eta^2}{f^2}+\Pi^L_{h\eta} \frac{h^2}{f^2}\frac{\eta^2}{f^2}$ & &\\ 

\hline

\rule[-2ex]{0pt}{5.5ex} 
\multirow{2}{*}{$\rm NMCHM_{20_L-6_R}$}  & $\Pi^L_0+\Pi^L_1 \frac{h^2}{f^2}+\Pi^L_2 \frac{h^4}{f^4}$ & \multirow{2}{*}{$\Pi^R_0+\Pi^R_1 \frac{h^2}{f^2}+\Pi^R_\eta \frac{\eta^2}{f^2}$} & \multirow{2}{*}{$\frac{h}{f}\left(\Pi^{LR}_1+\Pi^{LR}_2 \frac{h^2}{f^2}+\Pi^{LR}_\eta\frac{\eta^2}{f^2}\right)$} \\ 
& $+\Pi^L_\eta \frac{\eta^2}{f^2}+\Pi^L_{h\eta} \frac{h^2}{f^2}\frac{\eta^2}{f^2}$ & & \\
\hline 

\rule[-2ex]{0pt}{5.5ex} 
\multirow{2}{*}{$\rm NMCHM_{20_L-15_R}$} & $\Pi^L_0+\Pi^L_1 \frac{h^2}{f^2}+\Pi^L_2 \frac{h^4}{f^4}$ & \multirow{2}{*}{$\Pi^R_0+\Pi^R_1 \frac{h^2}{f^2}$}  & \multirow{2}{*}{$\Pi^{LR}_1\frac{h}{f}\sqrt{1-\frac{h^2}{f^2}-\frac{\eta^2}{f^2}}$} \\ 
& $+\Pi^L_\eta \frac{\eta^2}{f^2}+\Pi^L_{h\eta} \frac{h^2}{f^2}\frac{\eta^2}{f^2}$ & & \\
\hline 

\rule[-2ex]{0pt}{5.5ex} 
\multirow{2}{*}{$\rm NMCHM_{20_L-20_R}$}  & $\Pi^L_0+\Pi^L_1 \frac{h^2}{f^2}+\Pi^L_2 \frac{h^4}{f^4}$ & \multirow{2}{*}{$\Pi^R_0+\Pi^R_1 \frac{h^2}{f^2}+\Pi^R_2 \frac{h^4}{f^4}$} & \multirow{2}{*}{$\frac{h}{f}\sqrt{1-\frac{h^2}{f^2}-\frac{\eta^2}{f^2}}\left(\Pi^{LR}_1+\Pi^{LR}_2 \frac{h^2}{f^2}\right)$} \\ 
& $+\Pi^L_\eta \frac{\eta^2}{f^2}+\Pi^L_{h\eta} \frac{h^2}{f^2}\frac{\eta^2}{f^2}$ & & \\
\hline
%\end{longtabu}
\caption{\small\it List of $\Pi$-functions for different representations of next-to-minimal model.}
\label{nmchm_table1}
%\end{longtable}
\end{longtabu}
%\end{table}
%\end{small}

% % % % % % % % % % % % % % % % % % % % % % % % % % % % % % % % % % % % % %

\section{Expressions for $\Delta_t$, $\delta_t$ and $\Delta^\eta_t$}
\label{delta}

In Table~\ref{exp_model_table2} we list the expressions for
$\Delta_t$ and $\delta_t$ for the models $\rm MCHM_{5_L-5_R}$, $\rm
MCHM_{14_L-14_R}$, $\rm MCHM_{14_L-5_R}$ and $\rm MCHM_{5_L-14_R}$,
respectively.
\begin{longtable}{|c|c|c|c|}
\hline

\rule[-2ex]{0pt}{5.5ex} Models & $\Delta_t$ & $\delta_t$ \\ 

\hline 

\rule[-2ex]{0pt}{5.5ex} $\rm MCHM_{5_L-5_R}$ & $-\frac{3}{2}$ & $-\left(\frac{\Pi^{L}_1}{\Pi^{L}_0}+\frac{\Pi^{R}_1}{\Pi^{R}_0}\right)$ \\ 

\hline 

\rule[-2ex]{0pt}{5.5ex} $\rm MCHM_{14_L-14_R}$ & $2\frac{\Pi^{LR}_2}{\Pi^{LR}_1}-\frac{3}{2}$ & $-\left(\frac{\Pi^{L}_1}{\Pi^{L}_0}+\frac{\Pi^{R}_1}{\Pi^{R}_0}\right)$ \\ 

\hline 

\rule[-2ex]{0pt}{5.5ex} $\rm MCHM_{14_L-5_R}$ & $2\frac{\Pi^{LR}_2}{\Pi^{LR}_1}-\frac{1}{2}$ & $-\left(\frac{\Pi^{L}_1}{\Pi^{L}_0}+\frac{\Pi^{R}_1}{\Pi^{R}_0}\right)$  \\ 

\hline

\rule[-2ex]{0pt}{5.5ex} $\rm MCHM_{5_L-14_R}$ & $2\frac{\Pi^{LR}_2}{\Pi^{LR}_1}-\frac{1}{2}$ & $-\left(\frac{\Pi^{L}_1}{\Pi^{L}_0}+\frac{\Pi^{R}_1}{\Pi^{R}_0}\right)$ \\ 

\hline 

\caption{\small\it Expressions for $\Delta_t$ and $\delta_t$ for different representations of SO(5) in which top quark is embedded.}
\label{exp_model_table2}
\end{longtable}
%\end{table}

We present the expressions for $(\Delta_t+\delta_t)$ and $\Delta^\eta_t$, as defined in Eq.~\eqref{nmchm_3}, in terms of the form factors, for different $\rm SO(6)$ representations in Table~\ref{nmchm_delta_table}. 
\begin{longtabu} to \linewidth {|P{2.95cm}|P{1.045cm}|P{10.825cm}|}
\hline \rule[-2ex]{0pt}{5.0ex} Models & \multicolumn{2}{c|}{Coupling Modifiers}\\ 

\hline \rule[-2ex]{0pt}{5.0ex} $\rm NMCHM_{6_L-1_R}$ & $\Delta_t+\delta_t$ & $- \left(\frac{\Pi^L_1}{\Pi^L_0}+\frac{1}{2}\right)$ \\ 
\cline{2-3} \rule[-2ex]{0pt}{5.0ex}
& $\Delta^\eta_t$ & $0$\\ 

\hline \rule[-2ex]{0pt}{5.0ex} $\rm NMCHM_{6_L-6_R}$ & $\Delta_t+\delta_t$ & \pbox{10.5cm}{$-\left(1+\frac{\Pi^R_\eta}{\Pi^R_0}\chi\right)^{-1}\left[\frac{\Pi^{L}_1}{\Pi^{L}_0}+\frac{\Pi^{R}_1}{\Pi^{R}_0}+\frac{3}{2}+\left(\frac{\Pi^{L}_1}{\Pi^{L}_0}\frac{\Pi^{R}_\eta}{\Pi^{R}_0}+\frac{1}{2}\frac{\Pi^{R}_\eta}{\Pi^{R}_0}\right)\chi\right]$} \\ 
\cline{2-3} \rule[-2ex]{0pt}{5.0ex}
& $\Delta^\eta_t$ & $\left(1+\frac{\Pi^R_\eta}{\Pi^R_0}\right)\left(1+\frac{\Pi^R_\eta}{\Pi^R_0}\chi\right)^{-1}\sqrt{\frac{\chi}{1-\chi}}$\\ 

\hline \rule[-2ex]{0pt}{5.0ex} $\rm NMCHM_{6_L-15_R}$ & $\Delta_t+\delta_t$ & $-\left[\frac{\Pi^{L}_1}{\Pi^{L}_0}+\frac{\Pi^{R}_1}{\Pi^{R}_0}+\frac{1}{2}\right]$ \\ 
\cline{2-3} \rule[-2ex]{0pt}{5.0ex}
& $\Delta^\eta_t$ & $0$ \\ 
\hline
\newpage
\hline \rule[-2ex]{0pt}{5.0ex} $\rm NMCHM_{6_L-20_R}$ & $\Delta_t+\delta_t$ & $\left[2\frac{\Pi^{LR}_2}{\Pi^{LR}_1}-\frac{\Pi^{L}_1}{\Pi^{L}_0}-\frac{\Pi^{R}_1}{\Pi^{R}_0}-\frac{1}{2}\right]$ \\ 
\cline{2-3} \rule[-2ex]{0pt}{5.5ex}
& $\Delta^\eta_t$ & $0$ \\ 

\hline \rule[-2ex]{0pt}{5.0ex} $\rm NMCHM_{15_L-6_R}$ & $\Delta_t+\delta_t$ & \pbox{10.5cm}{$-\left(1+\frac{\Pi^L_\eta}{\Pi^L_0}\chi\right)^{-1}\left(1+\frac{\Pi^R_\eta}{\Pi^R_0}\chi\right)^{-1}\left[\frac{\Pi^{L}_1}{\Pi^{L}_0}+\frac{\Pi^{R}_1}{\Pi^{R}_0}+\frac{1}{2}\right.$\\$\left.-\left(\frac{1}{2}\frac{\Pi^{L}_\eta}{\Pi^{L}_0}+\frac{1}{2}\frac{\Pi^{R}_\eta}{\Pi^{R}_0}-\frac{\Pi^{L}_1}{\Pi^{L}_0}\frac{\Pi^{R}_\eta}{\Pi^{R}_0}-\frac{\Pi^{R}_1}{\Pi^{R}_0}\frac{\Pi^{L}_\eta}{\Pi^{L}_0}\right)\chi-\frac{3}{2}\frac{\Pi^{L}_\eta}{\Pi^{L}_0}\frac{\Pi^{R}_\eta}{\Pi^{R}_0}\chi^2\right]$} \\ 
\cline{2-3} \rule[-2ex]{0pt}{5.0ex}
& $\Delta^\eta_t$ & \pbox{12.5cm}{$(1-\chi)\left(1+\frac{\Pi^L_\eta}{\Pi^L_0}\chi\right)^{-1}\left(1+\frac{\Pi^R_\eta}{\Pi^R_0}\chi\right)^{-1}\left[\frac{\Pi^{L}_\eta}{\Pi^{L}_0}+\frac{\Pi^{R}_\eta}{\Pi^{R}_0}+2\frac{\Pi^{L}_\eta}{\Pi^{L}_0}\frac{\Pi^{R}_\eta}{\Pi^{R}_0}\chi\right]\sqrt{\frac{\chi}{1-\chi}}$} \\ 

\hline \rule[-2ex]{0pt}{5.0ex} $\rm NMCHM_{15_L-15_R}$ & $\Delta_t+\delta_t$ & $-\left(1+\frac{\Pi^L_\eta}{\Pi^L_0}\chi\right)^{-1}\left[\frac{\Pi^L_1}{\Pi^L_0}+\frac{\Pi^R_1}{\Pi^R_0}+\frac{3}{2}+\left(\frac{1}{2}\frac{\Pi^L_\eta}{\Pi^L_0}+\frac{\Pi^L_{\eta}}{\Pi^L_0}\frac{\Pi^R_{1}}{\Pi^R_0}\right)\chi\right]$ \\ 
\cline{2-3} \rule[-2ex]{0pt}{5.0ex}
& $\Delta^\eta_t$ & $\left(1+\frac{\Pi^L_\eta}{\Pi^L_0}\right)\left(1+\frac{\Pi^L_\eta}{\Pi^L_0}\chi\right)^{-1}\sqrt{\frac{\chi}{1-\chi}}$\\ 

\hline \rule[-2ex]{0pt}{5.0ex} $\rm NMCHM_{15_L-20_R}$ & $\Delta_t+\delta_t$ & $-\left(1+\frac{\Pi^L_\eta}{\Pi^L_0}\chi\right)^{-1}\left[\frac{\Pi^L_1}{\Pi^L_0}+\frac{\Pi^R_1}{\Pi^R_0}+\frac{3}{2}+\left(\frac{1}{2}\frac{\Pi^L_\eta}{\Pi^L_0}+\frac{\Pi^L_{\eta}}{\Pi^L_0}\frac{\Pi^R_{1}}{\Pi^R_0}\right)\chi\right]$ \\ 
\cline{2-3} \rule[-2ex]{0pt}{5.0ex}
& $\Delta^\eta_t$ & $\left(1+\frac{\Pi^L_\eta}{\Pi^L_0}\right)\left(1+\frac{\Pi^L_\eta}{\Pi^L_0}\chi\right)^{-1}\sqrt{\frac{\chi}{1-\chi}}$\\ 

\hline \rule[-2ex]{0pt}{5.0ex} $\rm NMCHM_{20_L-1_R}$ & $\Delta_t+\delta_t$ & $-\left(1+\frac{\Pi^L_\eta}{\Pi^L_0}\chi\right)^{-1}\left[\frac{\Pi^L_1}{\Pi^L_0}+\frac{3}{2}+\left(\frac{1}{2}\frac{\Pi^L_\eta}{\Pi^L_0}+\frac{\Pi^L_{h\eta}}{\Pi^L_0}\right)\chi\right]$ \\ 
\cline{2-3} \rule[-2ex]{0pt}{5.0ex}
& $\Delta^\eta_t$ & $\left(1+\frac{\Pi^L_\eta}{\Pi^L_0}\right)\left(1+\frac{\Pi^L_\eta}{\Pi^L_0}\chi\right)^{-1}\sqrt{\frac{\chi}{1-\chi}}$\\ 

\hline \rule[-2ex]{0pt}{5.0ex} $\rm NMCHM_{20_L-6_R}$ & $\Delta_t+\delta_t$ & \pbox{13.5cm}{$\left(1+\frac{\Pi^L_\eta}{\Pi^L_0}\chi\right)^{-1}\left(1+\frac{\Pi^R_\eta}{\Pi^R_0}\chi\right)^{-1}\left(1+\frac{\Pi^{LR}_\eta}{\Pi^{LR}_1}\chi\right)^{-1}\left[2\frac{\Pi^{LR}_2}{\Pi^{LR}_1}-\frac{\Pi^{L}_1}{\Pi^{L}_0}-\frac{\Pi^{R}_1}{\Pi^{R}_0}-\frac{1}{2}\right.$\\$\left.+\left(\frac{1}{2}\frac{\Pi^{L}_\eta}{\Pi^{L}_0}+\frac{1}{2}\frac{\Pi^{R}_\eta}{\Pi^{R}_0}-\frac{5}{2}\frac{\Pi^{LR}_\eta}{\Pi^{LR}_1}-\frac{\Pi^{L}_{h\eta}}{\Pi^{L}_0}+2\frac{\Pi^{LR}_2}{\Pi^{LR}_1}\frac{\Pi^{L}_\eta}{\Pi^{L}_0}+2\frac{\Pi^{LR}_2}{\Pi^{LR}_1}\frac{\Pi^{R}_\eta}{\Pi^{R}_0}-\frac{\Pi^{L}_1}{\Pi^{L}_0}\frac{\Pi^{R}_\eta}{\Pi^{R}_0}\right.\right.$\\$\left.\left.-\frac{\Pi^{R}_1}{\Pi^{R}_0}\frac{\Pi^{L}_\eta}{\Pi^{L}_0}-\frac{\Pi^{L}_1}{\Pi^{L}_0}\frac{\Pi^{LR}_\eta}{\Pi^{LR}_1}-\frac{\Pi^{R}_1}{\Pi^{R}_0}\frac{\Pi^{LR}_\eta}{\Pi^{LR}_1}\right)\chi+\left(\frac{3}{2}\frac{\Pi^{L}_\eta}{\Pi^{L}_0}\frac{\Pi^{R}_\eta}{\Pi^{R}_0}-\frac{3}{2}\frac{\Pi^{L}_\eta}{\Pi^{L}_0}\frac{\Pi^{LR}_\eta}{\Pi^{LR}_1}-\frac{3}{2}\frac{\Pi^{R}_\eta}{\Pi^{R}_0}\frac{\Pi^{LR}_\eta}{\Pi^{LR}_1}\right.\right.$\\$\left.\left.-\frac{\Pi^{L}_{h\eta}}{\Pi^{L}_0}\frac{\Pi^{R}_\eta}{\Pi^{R}_0}-\frac{\Pi^{L}_{h\eta}}{\Pi^{L}_0}\frac{\Pi^{LR}_\eta}{\Pi^{LR}_1}+2\frac{\Pi^{LR}_2}{\Pi^{LR}_1}\frac{\Pi^{L}_\eta}{\Pi^{L}_0}\frac{\Pi^{R}_\eta}{\Pi^{R}_0}-\frac{\Pi^{R}_1}{\Pi^{R}_0}\frac{\Pi^{L}_\eta}{\Pi^{L}_0}\frac{\Pi^{LR}_\eta}{\Pi^{LR}_1}-\frac{\Pi^{L}_1}{\Pi^{L}_0}\frac{\Pi^{R}_\eta}{\Pi^{R}_0}\frac{\Pi^{LR}_\eta}{\Pi^{LR}_1}\right)\chi^2\right.$\\$\left.+\left(-\frac{1}{2}\frac{\Pi^{L}_\eta}{\Pi^{L}_0}\frac{\Pi^{R}_\eta}{\Pi^{R}_0}\frac{\Pi^{LR}_\eta}{\Pi^{LR}_1}-\frac{\Pi^{L}_{h\eta}}{\Pi^{L}_0}\frac{\Pi^{R}_\eta}{\Pi^{R}_0}\frac{\Pi^{LR}_\eta}{\Pi^{LR}_1}\right)\chi^3\right]$} \\ 
\cline{2-3} \rule[-2ex]{0pt}{5.0ex}
& $\Delta^\eta_t$ & \pbox{13.5cm}{$-(1-\chi)\left(1+\frac{\Pi^L_\eta}{\Pi^L_0}\chi\right)^{-1}\left(1+\frac{\Pi^R_\eta}{\Pi^R_0}\chi\right)^{-1}\left(1+\frac{\Pi^{LR}_\eta}{\Pi^{LR}_1}\chi\right)^{-1}$\\$\left[2\frac{\Pi^{LR}_\eta}{\Pi^{LR}_1}-\frac{\Pi^{L}_\eta}{\Pi^{L}_0}-\frac{\Pi^{R}_\eta}{\Pi^{R}_0}+\left(\frac{\Pi^{L}_\eta}{\Pi^{L}_0}\frac{\Pi^{LR}_\eta}{\Pi^{LR}_1}+\frac{\Pi^{R}_\eta}{\Pi^{R}_0}\frac{\Pi^{LR}_\eta}{\Pi^{LR}_1}-2\frac{\Pi^{L}_\eta}{\Pi^{L}_0}\frac{\Pi^{R}_\eta}{\Pi^{R}_0}\right)\chi\right]\sqrt{\frac{\chi}{1-\chi}}$} \\ 

\hline \rule[-2ex]{0pt}{5.0ex} $\rm NMCHM_{20_L-15_ R}$ & $\Delta_t+\delta_t$ & $-\left(1+\frac{\Pi^L_\eta}{\Pi^L_0}\chi\right)^{-1}\left[\frac{\Pi^L_1}{\Pi^L_0}+\frac{\Pi^R_1}{\Pi^R_0}+\frac{3}{2}+\left(\frac{1}{2}\frac{\Pi^L_\eta}{\Pi^L_0}+\frac{\Pi^L_{h\eta}}{\Pi^L_0}+\frac{\Pi^L_{\eta}}{\Pi^L_0}\frac{\Pi^R_{1}}{\Pi^R_0}\right)\chi\right]$ \\ 
\cline{2-3} \rule[-2ex]{0pt}{5.0ex}
& $\Delta^\eta_t$ & $\left(1+\frac{\Pi^L_\eta}{\Pi^L_0}\right)\left(1+\frac{\Pi^L_\eta}{\Pi^L_0}\chi\right)^{-1}\sqrt{\frac{\chi}{1-\chi}}$\\ 

\hline \rule[-2ex]{0pt}{5.0ex} $\rm NMCHM_{20_L-20_R}$ & $\Delta_t+\delta_t$ & \pbox{12.6cm}{$\left(1+\frac{\Pi^L_\eta}{\Pi^L_0}\chi\right)^{-1}\left[2\frac{\Pi^{LR}_2}{\Pi^{LR}_1}-\frac{\Pi^{L}_1}{\Pi^{L}_0}-\frac{\Pi^{R}_1}{\Pi^{R}_0}-\frac{3}{2}\right.$\\$\left.+\left(2\frac{\Pi^{LR}_2}{\Pi^{LR}_1}\frac{\Pi^{L}_\eta}{\Pi^{L}_0}-\frac{\Pi^{R}_1}{\Pi^{R}_0}\frac{\Pi^{L}_\eta}{\Pi^{L}_0}-\frac{1}{2}\frac{\Pi^{L}_\eta}{\Pi^{L}_0}-\frac{\Pi^{L}_{h\eta}}{\Pi^{L}_0}\right)\chi\right]$} \\ 
\cline{2-3} \rule[-2ex]{0pt}{5.0ex}
& $\Delta^\eta_t$ & $\left(1+\frac{\Pi^L_\eta}{\Pi^L_0}\right)\left(1+\frac{\Pi^L_\eta}{\Pi^L_0}\chi\right)^{-1}\sqrt{\frac{\chi}{1-\chi}}$ \\ 

\hline

\caption{\small\it Expressions for $(\Delta_t+\delta_t)$ and $\Delta^\eta_t$ for different representations of $\rm SO(6)$ in which the top quark is embedded.}
\label{nmchm_delta_table}
\end{longtabu}

% % % % % % % % % % % % % % % % % % % % % % % % % % % % % % % % % % % % % %

\begin{center}
\rule{0.5\textwidth}{1pt}
\end{center}

% % % % % % % % % % % % % % % % % % % % % % % % % % % % % % % % % % % % %  

% % % % % % % % % % % % % % References % % % % % % % % % % % % % % % % % % 
 
\bibliographystyle{JHEP}
\bibliography{CompYukawa}

\end{document}